# Infrared multiple-angle incidence resolution spectrometry for vapor-deposited amorphous water

Takumi Nagasawa, Naoki Numadate, and Tetsuya Hama[*]

Komaba Institute for Science and Department of Basic Science, The University of Tokyo, Meguro, Tokyo 153-8902, Japan

[*]Author to whom correspondence should be addressed: hamatetsuya@g.ecc.u-tokyo.ac.jp

**Abstract**

Infrared (IR) multiple-angle incidence resolution spectrometry (IR-MAIRS) is a recently developed spectroscopic technique that combines oblique incidence transmission measurements and chemometrics (multivariate analysis) to obtain both pure in-plane (IP) and out-of-plane (OP) vibration spectra for a thin sample. IR-MAIRS is established for analyzing the molecular orientation of organic thin films at atmospheric pressure, but it should also be powerful for the structural characterization of vapor-deposited thin samples prepared in a vacuum. The application of IR-MAIRS to vapor-deposited amorphous water is particularly interesting in the fields of physical and interstellar chemistry, because it is a representative model material for interstellar icy dust grains. We recently developed an experimental setup for *in situ* IR-MAIRS under low-temperature, ultrahigh-vacuum conditions, which thus facilitates measurements of interstellar ice analogues such as vapor-deposited amorphous water. This review considers the theoretical framework of IR-MAIRS and our recent experimental results for vapor-deposited amorphous water. We present spectroscopic signatures for the perpendicular orientation of dangling OH bonds for three-coordinated water molecules at the surface of amorphous water at 90 K. The absolute absorption cross-section of the three-coordinated dangling OH bonds is quantitatively measured as $1.0 \pm 0.2 \times 10^{-18}$ cm$^2$ at 3696 cm$^{-1}$. As IR-MAIRS can essentially be conducted using only a Fourier-transform IR spectrometer and an angle-controllable linear polarizer, it is a useful, low-cost, and simple spectroscopic technique for studying laboratory analogues of interstellar ices including vapor-deposited amorphous water.

## 1. Introduction

Water ($H_2O$) is ubiquitous throughout the universe because its constituent hydrogen (H) and oxygen (O) are the most and third-most abundant elements, respectively.[1–6] Gaseous $H_2O$ molecules have been found in star- and planet-forming regions in interstellar space and in the atmospheres of planets both within and beyond our solar system.[6] Low-temperature (<50 K) interstellar regions such as dense molecular clouds, the birthplaces of stars, contain $H_2O$ mainly as a metastable amorphous solid covering the surfaces of interstellar dust grains (Fig. 1).[5] This icy dust is a dominant solid component of dense molecular clouds and the outer cold parts of protoplanetary disks, and has played a critical role as a building block for the formation of, for example, the gas and icy giants and comets in the early solar system.[7] Studying the physics and chemistry of interstellar ice thus helps us to understand the evolution of stars and planetary systems, including our solar system.[5–10]

Interstellar ices are directly detected through their molecular vibrational transitions at infrared (IR) wavelengths.[11] For example, Fig. 2 shows the IR spectrum obtained by observation of a low-mass protostar Elias 29.[12] The clear vibrational absorption bands of $H_2O$ ice at 3.1, 4.5, 6.0, and 11 μm are assigned respectively to OH-stretching, combination, OH-bending, and libration. Absorption bands from solid carbon monoxide (CO), carbon dioxide ($CO_2$), and silicates are also observed. This interstellar IR spectrum indicates that the protostar is surrounded by icy dust grains with silicate cores (Fig. 1). The water ice on these grains is predominantly amorphous, given the lack of a sharp OH-stretching feature around 3200 cm$^{-1}$ (3.1 μm) that is characteristic of four-coordinated molecules in crystalline ice.[5,6]

Table 1 summarizes the chemical compositions of the icy dust grains observed in the low-mass protostar Elias 29 (Fig. 2)[12] along with equivalent abundances of cometary volatiles in our solar system,[13] showing the qualitative good agreement between them. To identify the IR absorption bands in astronomical observations, derive the column densities (molecules cm$^{-2}$) of the observed molecules, and determine the upper limits of abundances for undetected molecules requires laboratory measurement of reference IR spectra of interstellar ice analogues and their optical constants (e.g., absorption cross-sections) for each IR band.[14,15] Astrochemical laboratory studies of interstellar ice analogues (e.g., molecular solids such as $H_2O$, $CO$, $CO_2$, $NH_3$, $CH_4$, and $CH_3OH$) usually employ samples prepared by vapor deposition on a cold substrate. IR measurements of these ices are generally performed with a Fourier-transform (FT)-IR spectrometer in normal-incidence transmission geometry or reflection geometry at grazing incidence using a metallic substrate (reflection–absorption measurement), as shown in Fig. 3. The Beer–Lambert law is often used to give a sample's absorbance ($A$) as a function of the imaginary part of its complex refractive index (also called the extinction coefficient), ($n + ik$):

$$A = -\log_{10} \frac{S^s}{S^b} = \frac{4\pi dk}{\lambda \ln 10}, (1)$$

where $S$ is the intensity of the transmitted (or reflected) IR light of wavelength $\lambda$ (μm) (i.e., single-beam spectrum); the superscripts $s$ and $b$ indicate background and sample measurements, respectively;[16] and $d$ is the sample thickness for transmission measurements or the optical path length in the sample for reflection–absorption measurements. However, this law holds only for a bulk sample whose thickness is greater than the IR wavelength ($d \geq \lambda$), because no optical interface (e.g., vacuum–sample–substrate interfaces) is considered in its derivation from Maxwell's equations.[16] However, the thickness of vapor-deposited amorphous water at low temperature in a vacuum is typically less than hundreds of nanometers, to prevent saturation of the absorption by the 3.1 μm (OH-stretching) band: i.e., $d \ll \lambda$.[17,18] In such cases, the thin sample's absorbance is induced by the electric field at optical interfaces (vacuum–sample–substrate interfaces), which is different from the far field.[16]

Many articles and textbooks have considered the electromagnetic theoretical treatment of IR spectroscopy.[16,19–23] The key approximations to derive analytical expressions for the absorbance of a thin sample are the thin-film approximation (eq. 2) and the weak-absorption approximation (eq. 3):

$$\frac{d}{\lambda} \ll 1, (2)$$

$$\frac{\Delta S}{S^b} \ll 1, (3)$$

where $\Delta S = S^b - S^s$. When these two approximations are satisfied, the analytical expression for the thin sample's absorbance is largely simplified.[16,19] For example, Itoh et al. deduced eqs. 4, 5, and 6 to describe the absorbance of a thin sample ($A_{\text{thin}}^{\theta=0°}$) for normal-incidence measurement of a typical five-layer, symmetric double-sided sample system (vacuum/thin sample/IR-transparent substrate/thin sample/vacuum; see Fig. 3A):[19]

$$A_{\text{thin}}^{\theta=0°} = -\log_{10}\frac{S^s}{S^b} = \frac{8\pi da}{\lambda \ln 10} f(\varepsilon)_{\text{TO}}, (4)$$

$$a = \frac{1}{n_v + n_s} + \left(\frac{n_v - n_s}{n_v + n_s}\right)^4 \left\{1 - \left(\frac{n_v - n_s}{n_v + n_s}\right)^4\right\}^{-1} \left(\frac{2n_s}{n_s^2 - n_v^2}\right), (5)$$

$$f(\varepsilon)_{\text{TO}} = \text{Im}[\overline{\varepsilon_{xy}}] = \text{Im}[(n_{xy} + ik_{xy})^2] = 2n_{xy}k_{xy}, (6)$$

where $n_v$ and $n_s$ are the refractive index of the vacuum and the substrate, respectively. The coefficient $a$ is expressed in terms of $n_v$ and $n_s$, and results from optical interface effects. The transverse optic energy-loss function of the thin sample, $f(\varepsilon)_{\text{TO}}$, is expressed using the *x*- and *y*-components (surface-parallel components) of the complex permittivity ($\overline{\varepsilon_{xy}}$) or complex refractive index ($n_{xy} + ik_{xy}$) of the thin sample, where $\bar{\varepsilon} = (n + ik)^2$ (see Fig. 3 for the definition of the *x*- and *y*-directions). In this article, a uniaxial system ($k_x = k_y = k_{xy}$) is assumed for simplicity. Eqs. 4, 5, and 6 show that $A_{\text{thin}}^{\theta=0°}$ is influenced by $n_v$ and $n_s$, despite $A_{\text{thin}}^{\theta=0°}$ being obtained by calculating the ratio between $S^s$ (sample and substrate) and $S^b$ (substrate only). Furthermore, the electric field of the IR light is parallel to the substrate for normal-incidence transmission measurements. Consequently, only surface-parallel (in-plane, IP) vibration is observed in the IR spectrum of the thin sample, whose band shape is expressed as $f(\varepsilon)_{\text{TO}}$.[16,19,20,24]

For reflection–absorption measurements (Fig. 3B), both the real and imaginary parts of the complex permittivity of the metal substrate ($\bar{\varepsilon}_s$) are significantly larger in the IR range than those of the thin sample ($\bar{\varepsilon}$) and vacuum ($\varepsilon_v$, real value); i.e., $|\varepsilon_s| \gg |\varepsilon|$ and $|\varepsilon_s| \gg \varepsilon_v$. Reflection–absorbance using p-polarized IR light ($A_p^{\text{RA}}$) can be simply expressed

using eqs. 7 and 8, when $|\varepsilon_s| \gg \varepsilon_v \tan^2\theta$ is satisfied; i.e., the angle of incidence $\theta$ is less than approximately 85°:[16]

$$A_p^{RA} = -\log_{10}\frac{S^s}{S^b} = \frac{8\pi d}{\lambda \ln 10}\frac{n_v^3 \sin^2\theta}{\cos\theta} f(\varepsilon)_{LO}, (7)$$

$$f(\varepsilon)_{LO} = \text{Im}\left[-\frac{1}{\overline{\varepsilon_z}}\right] = \text{Im}\left[\frac{1}{(n_z + ik_z)^2}\right] = \frac{2n_z k_z}{(n_z^2 + k_z^2)^2}, (8)$$

where $f(\varepsilon)_{LO}$ is the longitudinal optic energy-loss function of the thin sample, which is expressed using the z-component (surface-perpendicular component) of the complex permittivity ($\overline{\varepsilon_z}$) or complex refractive index ($n_z + ik_z$) of the thin sample (Fig. 3). Reflection–absorption measurements have the electric field of the IR light perpendicular to the metallic substrate; therefore, only the surface-perpendicular (OP) vibration is observed in the IR spectrum of the thin sample, and the band shape is given by $f(\varepsilon)_{LO}$.[16,20,24] For reference, negligible reflection–absorbance is induced by s-polarized IR light. A reflection–absorption measurement using unpolarized light gives about half the absorbance ($\frac{A_p^{RA}}{2}$) when the intensities of p- and s-polarized light can be regarded as identical.[16]

Eqs. 4–8 show that the analytical expressions for $A_{thin}^{\theta=0°}$ and $A_p^{RA}$ are different, meaning that a given sample would show different band shapes and intensities in its IR transmission and reflection–absorption spectra. For example, Fig. 4A and B shows IR normal-incidence transmission and reflection–absorption spectra of vapor-deposited amorphous water at 10 and 11 K, respectively, prepared at a similar deposition pressure of $1 \times 10^{-5}$ Pa. Normal-incidence measurements use a Si substrate, while reflection–absorption measurements use Al. Data for normal-incidence transmission are measured in our laboratory (see section 2.1. for experimental details), and those for reflection–absorption spectra are adapted from ref.[25] Overall, both the normal-incidence transmission and reflection–absorption spectra have similar band shapes and intensities. However, detailed inspection of the OH stretching band shows clearly different peak wavenumbers between the normal-incidence transmission (3309 cm$^{-1}$) and reflection–absorption (3418 cm$^{-1}$) (Fig. 5A). The peak wavenumber of the OH-stretching band has often been suggested to correlate with the hydrogen-bond strength.[16] However, such a large peak shift would not be caused by a change in strength of the hydrogen-bonding between amorphous water molecules that originated from different water−substrate (Si and Al) interactions, considering the similar substrate temperatures and deposition conditions. This peak shift should rather be attributed to the transverse optic (TO)–longitudinal optic (LO) splitting, because the normal-incidence transmission and reflection–absorption spectra are expressed with $f(\varepsilon)_{\mathrm{TO}}$ and $f(\varepsilon)_{\mathrm{LO}}$, respectively. TO–LO splitting is induced by the dispersion of the real values (*n*) of the complex refractive index of amorphous water induced by the large imaginary (*k*) values; namely, strong IR absorption in the region of OH stretching vibration (Fig. 5B).[14,18,26] The band shapes of the normal-incidence and reflection–absorption spectra are consistent with those for $f(\varepsilon)_{\mathrm{TO}}$ and $f(\varepsilon)_{\mathrm{LO}}$ calculated using *n* and *k* values reported for amorphous water at 10[14] and 15 K,[18] as shown in Fig. 5A and C. These examples demonstrate that a careful consideration based on $f(\varepsilon)_{\mathrm{TO}}$ and $f(\varepsilon)_{\mathrm{LO}}$ is necessary for the proper interpretation of measurements of a thin ice sample, especially when comparing $A_{\mathrm{thin}}^{\theta=0°}$ and $A_{\mathrm{p}}^{\mathrm{RA}}$.

In addition, eqs. 4–8 show surface selection rules; i.e., surface-parallel and surface-perpendicular components of a transition moment are selectively observed in the normal-incidence transmission and reflection–absorption spectra, respectively.[16,22,23] These rules hinder both the analysis of molecular orientation and the quantification of the column density of the molecules in a thin sample, because the IR band intensities ($A_{\text{thin}}^{\theta=0°}$ and $A_{\text{p}}^{\text{RA}}$) of a thin sample are influenced not only by the column density, but also by the molecular orientation.[24,27]

To overcome these limitations, Hasegawa developed a novel IR measurement method in 2002 called multiple-angle incidence resolution spectrometry (IR-MAIRS) (see section 2 for details).[28] This method combines oblique incidence transmission measurements and multivariate analysis to retrieve both pure IP and OP vibration spectra (i.e., $f(\varepsilon)_{\text{TO}}$ and $f(\varepsilon)_{\text{LO}}$) for a thin sample. Following refinement of the measurement and analytical procedures by Hasegawa and co-workers,[19,24,28–32] IR-MAIRS is now widely used for the molecular orientation analysis of organic thin films.[24] Recently, we developed an experimental setup for *in situ* IR-MAIRS under low-temperature, ultrahigh-vacuum conditions.[33,34] This article reviews our recent results for vapor-deposited amorphous water using low-temperature, ultrahigh-vacuum IR-MAIRS. We describe the experimental apparatus and the theoretical framework in detail, and present some specific experimental results for vapor-deposited amorphous water, with emphasis on the molecular orientation and absorption cross-section of dangling OH bonds; i.e., the free OH stretching modes of three-coordinated (one H donor and two H acceptors) and two-coordinated (one H donor and one H acceptor) molecules located at the ice surface (Fig. 6).[35–38] We demonstrate that more detailed information about the physicochemical properties of vapor-deposited amorphous water can be obtained by using IR-MAIRS than by using conventional normal-incidence transmission and reflection−absorption measurements.

## 2. Infrared multiple-angle incidence resolution spectrometry (IR-MAIRS)

### 2.1. IR-MAIRS measurements

Our experimental apparatus for low-temperature, ultrahigh-vacuum IR-MAIRS comprised a vacuum sample chamber and a FT-IR spectrometer (Fig. 7). The vacuum sample chamber was evacuated to ultrahigh-vacuum conditions (base pressure $10^{-8}$ Pa at room temperature) using a turbo molecular pump (STP-451, 480 L s$^{-1}$ for N$_2$, Edwards). A Si(111) substrate (40 × 40 × 1 mm, Pier Optics) was mounted on a copper sample holder without removing the native oxide on the Si surface. The Si substrate and copper sample holder were connected with indium solder by ultrasonic soldering under a nitrogen atmosphere to ensure good thermal contact for cooling the Si substrate to 6 K (Fig. 7).[33] The copper sample holder was connected to the cold head of a closed-cycle He refrigerator (RDK-101D, Sumitomo Heavy Industries) and installed in the vacuum sample chamber using a bore-through rotary feedthrough (KRP-152, Kitano Seiki). The inner wall of the vacuum chamber was coated with graphite particles (Aerodag G, Henkel Japan) to reduce the noise caused by scattering of IR light in the vacuum chamber. The Si substrate in the chamber was installed in the sample compartment of the FT-IR spectrometer (Nicolet iS50, Thermo Fisher Scientific) in transmission geometry across two 3 mm-thick ZnSe windows in the chamber. As the He refrigerator was freely rotatable by the bore-through rotary feedthrough, the angle of incidence of the IR beam ($\theta$) could be varied by rotating the Si substrate. The temperature of the substrate was measured using a Si diode sensor (DT-670, Lakeshore) placed in the copper sample holder and controlled with accuracies of ±0.2 K at 10 K and ±0.01 K at 90 K using a temperature controller (Model 325, Lakeshore) and a 40 W ceramic heater (MC1010, Sakaguchi E.H Voc Corp.).

Purified $H_2O$ (resistivity ≥ 18.2 MΩ cm at 298 K) from a Millipore Milli-Q water purification system was mixed with 2.0 wt.% (1.8 mol%) $D_2O$ (deuteration degree > 99.9%, Merck) to obtain a water sample containing about 3.5 mol% HDO with a negligible amount of $D_2O$. The water sample containing HDO facilitates measurement of the OD stretching vibration of HDO molecules in the bulk ice decoupled from intramolecular and intermolecular OH stretching vibrations.[39–45] The decoupled OD stretching band is useful, because its peak wavenumber is sensitive to the local lattice structure (oxygen–oxygen distance) in the ice.[25,44–57] Water was first degassed by several freeze–pump–thaw cycles, and then background vapor deposited at $2.2^{+1.3}_{-0.6} \times 10^{-6}$ Pa to form amorphous water, which corresponds to a flux of $8.0^{+4.4}_{-2.1} \times 10^{12}$ molecules cm$^{-2}$ s$^{-1}$. The pressure was measured using a cold cathode gauge (423 I-Mags® Cold Cathode Vacuum Sensor, MKS) and calculated using the gas correction factor for $H_2O$ (1.25 ± 0.44).[58] Hence, for example, amorphous water formed by vapor deposition for 32, 64, and 128 min corresponds to exposures of $1.5^{+0.8}_{-0.4} \times 10^{16}$, $3.1^{+1.7}_{-0.8} \times 10^{16}$, and $6.2^{+3.3}_{-1.6} \times 10^{16}$ molecules cm$^{-2}$, respectively. For reference, the lattice parameters of hexagonal ice indicate about $1.0 \times 10^{15}$ molecules cm$^{-2}$ on the surface.[59] Methanol deposition experiments used

CH$_3$OH (99%; Nacalai Tesque). The deposition pressure was calculated as 2.2 ± 0.2 × 10$^{-8}$ Pa using the gas correction factor for CH$_3$OH (1.69–2.12).[60–62] The flux of CH$_3$OH was estimated as 7.8 ± 0.9 × 10$^{10}$ molecules cm$^{-2}$ s$^{-1}$.

There are currently two types of measurement system: pMAIRS[29] and MAIRS2.[32] In pMAIRS, IR spectra of a thin sample on a substrate (Si in this study) are collected with eight incident angles, $\theta$, of 9°–44° in 5° steps using p-polarized IR light.[29–31] In MAIRS2, the IR spectra of a thin sample on a substrate are collected at a fixed incident angle of $\theta = 45°$ with seven polarization angles ($\phi$) from s-polarization ($\phi = 0°$) to p-polarization ($\phi = 90°$) in 15° steps (Figs. 8 and 9). Compared with pMAIRS, which requires a low angle of incidence measurement ($\theta = 9°$), MAIRS2 has the advantage that noise components due to optical interference fringes from multiple reflections of IR light in the substrate do not appear in the absorption spectrum thanks to the oblique incident angle being fixed at $\theta = 45°$, which leads to a better signal-to-noise ratio than achieved by pMAIRS. We thus used MAIRS2 to obtain IP and OP spectra. Hereafter, MAIRS2 is simply referred to as IR-MAIRS.

For IR-MAIRS measurements, the FT-modulated IR light was passed through an angle-controllable linear polarizer comprising a ZnSe wire grid incorporated in the FT-IR spectrometer, and oblique incidence transmission measurements were taken at $\theta = 45°$. IR-MAIRS measurements at seven polarization angles from $\phi = 0°$ to 90° in 15° steps were performed using OMNIC software for the FT-IR spectrometer (Nicolet iS50).[63] A mercury cadmium telluride detector was used to detect the IR light, and the intensity of the transmitted IR light was measured as a single-beam by the FT-IR.[32,33] The accumulation number of the single-beam measurements was 1000 (615 s) for each polarization angle with a resolution of 4 cm$^{-1}$. The total measurement time was 4305 s (72 min) for one IR-MAIRS measurement.

The baselines of the IP and OP spectra in the figures in this article are corrected and offset for clarity. The IP and OP spectra's vertical axes (absorbance) given here are $A_{\text{IP}}^{\theta=0°}$ and $n^4 H A_{\text{OP}}^{\theta=0°}$, respectively, because they reflect $k_{xy}$ and $k_z$ with the common ordinate scale as $A_{\text{IP}}^{\theta=0°}/n^4 H A_{\text{OP}}^{\theta=0°} = k_{xy}/k_z$ (for details, see eqs. 54–62 in section 5).

## 2.2. IR-MAIRS analysis

Each IR spectrum measured by seven oblique incidence transmission measurements contains information about both the IP and OP vibration modes; i.e., it is a mixed IP and OP spectrum.[19,28,32] IR-MAIRS employs classical least-squares (CLS) modeling to decompose the mixed IP and OP spectra into two pure IP and OP spectra.[16] The intensity of the transmitted IR light measured as single-beam spectra ($\boldsymbol{S}_{\text{obs}}$) with polarization angle $\phi$ and incident angle $\theta$ (fixed at 45°) can be expressed as a linear combination of the IP and OP polarization components ($\boldsymbol{S}_{\text{IP}}$ and $\boldsymbol{S}_{\text{OP}}$, respectively) with weighting ratios ($r_{\text{IP}}$ and $r_{\text{OP}}$, respectively) and a nonlinear noise factor $U$ (related to, for example, reflected IR light):[16,28]

$$\boldsymbol{S}_{\text{obs}} = r_{\text{IP}}\boldsymbol{S}_{\text{IP}} + r_{\text{OP}}\boldsymbol{S}_{\text{OP}} + U. \quad (9)$$

The collection of the $j$th single-beam spectrum, $S_{\text{obs},j}$ ($j = 1, 2, \ldots, 7$), at a polarization angle of $\phi_j$ ($j = 1, 2, \ldots, 7$) forms the matrix, $S$, and the linear combination part can be expressed using CLS regression (eq. 10):

$$S = \begin{pmatrix} S_{\text{obs},1} \\ S_{\text{obs},2} \\ \vdots \\ S_{\text{obs},7} \end{pmatrix} = \begin{pmatrix} r_{\text{IP},1} & r_{\text{OP},1} \\ r_{\text{IP},2} & r_{\text{OP},2} \\ \vdots & \vdots \\ r_{\text{IP},7} & r_{\text{OP},7} \end{pmatrix} \begin{pmatrix} S_{\text{IP}} \\ S_{\text{OP}} \end{pmatrix} + \begin{pmatrix} U_1 \\ U_2 \\ \vdots \\ U_7 \end{pmatrix} \equiv R\sigma + U, (10)$$

where $R$ is a matrix of weighting coefficients of $r_{\text{IP},j}$ ($j = 1, 2\ldots 7$) and $r_{\text{OP},j}$ ($j = 1, 2\ldots 7$) for $S_{\text{IP}}$ and $S_{\text{OP}}$, respectively; $\sigma$ is the vector set of $S_{\text{IP}}$ and $S_{\text{OP}}$; and $U$ is a "garbage matrix" that receives the non-linear responses to $R$ (i.e., noise factors are rejected in the CLS calculation and pooled in $U$ as an error term). Therefore, when the weighting factor matrix, $R$, is available, $S_{\text{IP}}$ and $S_{\text{OP}}$ can be calculated as the least-squares solution of the CLS regression equation as follows. We first define $\xi(\sigma)$ as a function of vector set $\sigma$, indicating the square of distance between the true and observed data:

$$\xi(\sigma) \equiv \sum_j U_j^2 = U^T U. (11)$$

The main frame of the least-squares solution is to minimize the function $\xi(\sigma)$:

$$\min\{\xi(\sigma)\} = \min\{U^T U\} = \min\{(S - R\sigma)^T (S - R\sigma)\}. (12)$$

The condition of minimizing $\xi(\sigma)$ is as follows:

$$\frac{\partial \xi(\sigma)}{\partial \sigma} = -2R^T S + 2R^T R\sigma = 0. (13)$$

The solution in eq (13), $\hat{\sigma}$, exists only when the square matrix $R^T R$ has its inverse matrix $(R^T R)^{-1}$, which is given as eq. 14:

$$\hat{\sigma} = \begin{pmatrix} S_{IP} \\ S_{OP} \end{pmatrix} = (R^T R)^{-1} R^T \, S. \quad (14)$$

This least-squares solution corresponds to the optimal solution of eq (13). After obtaining the IP and OP single-beam spectra set $\hat{\sigma}$ ($S_{IP}$ and $S_{OP}$) for background and sample measurements, the IP and OP absorbance spectra ($A_{IP}$ and $A_{OP}$) are calculated as

$$A_{IP} = -\log_{10} \frac{S_{IP}^{sam}}{S_{IP}^{bg}}, \quad (15)$$

$$A_{OP} = -\log_{10} \frac{S_{OP}^{sam}}{S_{OP}^{bg}}, \quad (16)$$

where the superscripts bg and sam indicate background and sample measurements, respectively.

The above discussion shows that the weighting factor matrix, $R$, is the core part of MAIRS theory. The validity of this matrix is thus critical to MAIRS analysis, determining whether it can reasonably extract both pure IP and OP spectra and what these spectra physically mean. For IR-MAIRS (MAIRS2), Shioya et al. defined $R$ as

$$R = \begin{pmatrix} \gamma \cos^2 \phi_1 + \sin^2 \phi_1 (\cos^2 \theta + \sin^2 \theta \tan^2 \theta) & \sin^2 \phi_1 \tan^2 \theta \\ \gamma \cos^2 \phi_2 + \sin^2 \phi_2 (\cos^2 \theta + \sin^2 \theta \tan^2 \theta) & \sin^2 \phi_2 \tan^2 \theta \\ \vdots & \vdots \\ \gamma \cos^2 \phi_7 + \sin^2 \phi_7 (\cos^2 \theta + \sin^2 \theta \tan^2 \theta) & \sin^2 \phi_7 \tan^2 \theta \end{pmatrix}, \quad (17)$$

where $\gamma$ is a correction factor to account for the polarization dependence of the FT-IR spectrometer; i.e., the intensity ratio of s-polarized light ($S_{\text{blank}}^{\text{s}}$) to p-polarized light ($S_{\text{blank}}^{\text{p}}$) at a wavenumber when measured without a substrate (blank).[32] In practice, $\gamma$ can be easily calculated using the s- and p-polarized single-beam spectra of the background measurements with a substrate ($S_{\text{bg}}^{\text{s}}$ and $S_{\text{bg}}^{\text{p}}$, respectively) and the transmittance of the substrate for the s- and p-polarized light ($T_{\text{sub}}^{\text{s}}$ and $T_{\text{sub}}^{\text{p}}$, respectively):

$$\gamma \equiv \frac{S_{\text{blank}}^{\text{s}}}{S_{\text{blank}}^{\text{p}}} = \frac{S_{\text{bg}}^{\text{s}} / T_{\text{sub}}^{\text{s}}}{S_{\text{bg}}^{\text{p}} / T_{\text{sub}}^{\text{p}}}, \quad (18)$$

because $S_{\text{bg}}^{\text{s}} = S_{\text{blank}}^{\text{s}} T_{\text{sub}}^{\text{s}}$ and $S_{\text{bg}}^{\text{p}} = S_{\text{blank}}^{\text{p}} T_{\text{sub}}^{\text{p}}$. $T_{\text{sub}}^{\text{s}}$ and $T_{\text{sub}}^{\text{p}}$ at a given $\theta$ can be calculated using $n_{\text{v}}$ (1 for vacuum) and $n_{\text{s}}$ (3.41 for Si[23]):

$$T_{\text{sub}}^{\text{p}} = \frac{\left|\frac{4 n_{\text{v}} \cos\theta_3\, n_{\text{s}} \cos\theta}{(n_{\text{s}} \cos\theta + n_{\text{v}} \cos\theta_3)^2}\right|^2}{1 - \left|\frac{n_{\text{s}} \cos\theta - n_{\text{v}} \cos\theta_3}{n_{\text{v}} \cos\theta_3 + n_{\text{s}} \cos\theta}\right|^4}, \quad (19)$$

$$T_{\text{sub}}^{\text{s}} = \frac{\left|\frac{4 n_{\text{v}} \cos\theta\, n_{\text{s}} \cos\theta_3}{(n_{\text{v}} \cos\theta + n_{\text{s}} \cos\theta_3)^2}\right|^2}{1 - \left|\frac{n_{\text{v}} \cos\theta - n_{\text{s}} \cos\theta_3}{n_{\text{v}} \cos\theta + n_{\text{s}} \cos\theta_3}\right|^4}, \quad (20)$$

where $n_{\text{v}} \sin\theta = n_{\text{s}} \sin\theta_3$, and thus, $n_{\text{s}} \cos\theta_3 = \sqrt{n_{\text{s}}^2 - (n_{\text{v}} \sin\theta)^2}$.[16,19,23]

The derivation of the **R** matrix is briefly described elsewhere,[28,32] but a detailed explanation is helpful to understand how IR-MAIRS works and the physical meanings of the obtained $A_{IP}$ and $A_{OP}$. The next two subsections derive the $R$ matrix and analytical expressions for $A_{IP}$ and $A_{OP}$.

### 2.3. Derivation of the weighting factor matrix, $R$

The weighting factor matrix, $R$, can be derived by separating the intensity of the electric field vector into two pure IP and OP components in a five-layer system (Fig. 8). The electric field vector, $E$, of linear-polarized IR-light is expressed as

$$\boldsymbol{E} = \boldsymbol{i}|E_0|e^{i(\boldsymbol{k}\cdot\boldsymbol{r}-\omega t)}, (21)$$

where $|E_0|$ is the amplitude of vibration magnitude, $\boldsymbol{k}$ is a wave vector, $\omega$ is angular frequency, $\boldsymbol{r}$ is a position vector, and $t$ is time. The basis vector $\boldsymbol{i}$ corresponds to the polarization that can be written uniquely as a linear combination of the s-polarized and p-polarized basis vectors ($\boldsymbol{e_s}$ and $\boldsymbol{e_p}$, respectively). Therefore,

$$\boldsymbol{E} = [(\cos\phi)\boldsymbol{e_s} + (\sin\phi)\boldsymbol{e_p}]|E_0|e^{i(\boldsymbol{k}\cdot\boldsymbol{r}-\omega t)}. (22)$$

As the vector $\boldsymbol{e_s}$ is perpendicular to the plane of incidence, it is directly related to the IP component, specifically the $IP_y$ component (Fig. 8). Thus, we change the subscript of vector $\boldsymbol{e_s}$ to $\boldsymbol{e_{IP_y}}$:

$$\boldsymbol{E} = [(\cos\phi)\boldsymbol{e_{IP_y}} + (\sin\phi)\boldsymbol{e_p}]|E_0|e^{i(\boldsymbol{k}\cdot\boldsymbol{r}-\omega t)}. (23)$$

In contrast, the vector $\boldsymbol{e_p}$ includes both IP (*x*-) and OP (*z*-)components, and its decomposition requires other basic vectors $\boldsymbol{e_{IP_x}}$ and $\boldsymbol{e_{OP_z}}$ as follows (Fig. 10):

$$\boldsymbol{e_\mathrm{p}} = (\cos\theta)\boldsymbol{e}_{\mathrm{IP}_x} + (\sin\theta)\boldsymbol{e}_{\mathrm{OP}_z}. \quad (24)$$

Using eqs. 23 and 24, eq 22 can be rewritten as follows using the new coordinate related to the $\mathrm{IP}_x - \mathrm{IP}_y - \mathrm{OP}_z$ axis (Figs. 8 and 10):

$$\boldsymbol{E} = |\boldsymbol{E_0}|(\cos\phi)\boldsymbol{e}_{\mathrm{IP}_y}\, e^{i(\boldsymbol{k}\cdot\boldsymbol{r}-\omega t)} + |\boldsymbol{E_0}|(\sin\phi)(\cos\theta)\boldsymbol{e}_{\mathrm{IP}_x}\, e^{i(\boldsymbol{k}\cdot\boldsymbol{r}-\omega t)}$$
$$+ |\boldsymbol{E_0}|(\sin\phi)(\sin\theta)\boldsymbol{e}_{\mathrm{OP}_z}\, e^{i(\boldsymbol{k}\cdot\boldsymbol{r}-\omega t)}. \quad (25)$$

The first term represents the IP component of IR-light from the s-polarized state, and the second and third terms respectively represent the IP and OP components of the IR-light from the p-polarized state. As all three bases $(\boldsymbol{e}_{\mathrm{IP}_x}, \boldsymbol{e}_{\mathrm{IP}_y}, \boldsymbol{e}_{\mathrm{OP}_z})$ are orthogonal, the intensity of the electric field $E$ can be calculated as the inner product $\boldsymbol{E}\boldsymbol{E}^*$, where $\boldsymbol{E}^*$ is the complex conjugate vector of $\boldsymbol{E}$:

$$E = \boldsymbol{E}\boldsymbol{E}^* = |\boldsymbol{E_0}|^2\cos^2\phi + |\boldsymbol{E_0}|^2\sin^2\phi\cos^2\theta + |\boldsymbol{E_0}|^2\sin^2\phi\sin^2\theta = |\boldsymbol{E_0}|^2. \quad (26)$$

The simple consideration above gives a weighting factor matrix, $\boldsymbol{R}$, as follows:

$$\boldsymbol{R} = \begin{pmatrix} \cos^2\phi_1 + \sin^2\phi_1\cos^2\theta & \sin^2\phi_1\sin^2\theta \\ \cos^2\phi_2 + \sin^2\phi_2\cos^2\theta & \sin^2\phi_2\sin^2\theta \\ \vdots & \vdots \\ \cos^2\phi_7 + \sin^2\phi_7\cos^2\theta & \sin^2\phi_7\sin^2\theta \end{pmatrix}. \quad (27)$$

However, this matrix does not properly work as the weighting factor matrix for IR-MAIRS. Hasegawa's first IR-MAIRS paper showed that additional treatments are necessary for the $\sin\theta$ component from the p-polarization (i.e., $|\boldsymbol{E_0}|(\sin\phi)(\sin\theta)\boldsymbol{e}_{\mathrm{OP}_z}\, e^{i(\boldsymbol{k}\cdot\boldsymbol{r}-\omega t)}$), because the electric-field vector $\boldsymbol{e}_{\mathrm{OP}_z}$ propagates obliquely into the substrate at the incident angle of $\theta$, as shown in Fig. 11A.[28] Although this oblique incidence makes the situation difficult to conceptualize, the electric field vector can be regarded to move parallel to the surface when it propagates into the substrate during its transmission through the substrate (Fig. 11B). This means that there is an additional contribution to the IP component from the $\sin\theta$ component, which can be estimated as $|E_0|(\sin\phi)(\sin\theta)e_{OP_z} e^{i(k\cdot r-\omega t)} \times (\tan\theta)$ by geometrically considering the propagation distance. In a similar way, the OP component is estimated as $|E_0|(\sin\phi)(\sin\theta)e_{OP_z} e^{i(k\cdot r-\omega t)} \times \left(\frac{1}{\cos\theta}\right)$, because the interaction time of the $\sin\theta$ component with the film is elongated by $\frac{1}{\cos\theta}$. Consequently, we can rewrite the $\sin\theta$ component as follows:

$$|E_0|(\sin\phi)(\sin\theta)e_{OP_z} e^{i(k\cdot r-\omega t)}$$
$$= |E_0|(\sin\phi)(\sin\theta)(\tan\theta)e_{IP_x} e^{i(k\cdot r-\omega t)}$$
$$+ |E_0|(\sin\phi)(\tan\theta)e_{OP_z} e^{i(k\cdot r-\omega t)} . \quad (28)$$

Therefore, eq. 25 can be rewritten as

$$\boldsymbol{E} = |E_0|(\cos\phi)e_{IP_y} e^{i(k\cdot r-\omega t)} + |E_0|(\sin\phi)(\cos\theta)e_{IP_x} e^{i(k\cdot r-\omega t)}$$
$$+ |E_0|(\sin\phi)(\sin\theta)(\tan\theta)e_{IP_x} e^{i(k\cdot r-\omega t)}$$
$$+ |E_0|(\sin\phi)(\tan\theta)e_{OP_z} e^{i(k\cdot r-\omega t)} . \quad (29)$$

Now we define the term $E_{MAIRS}$ to represent the intensity of transported energy, which is separated into IP and OP components:

$$E_{MAIRS} = |E_0|^2 \cos^2\phi + |E_0|^2\{\sin^2\phi(\cos^2\theta + \sin^2\theta\tan^2\theta)\}$$
$$+ |E_0|^2 \sin^2\phi \tan^2\theta . \quad (30)$$

Therefore, the modified weighting factor matrix, $\mathbf{R}$, can be expressed as

$$\mathbf{R} = \begin{pmatrix} \cos^2\phi_1 + \sin^2\phi_1(\cos^2\theta + \sin^2\theta\tan^2\theta) & \sin^2\phi_1\tan^2\theta \\ \cos^2\phi_2 + \sin^2\phi_2(\cos^2\theta + \sin^2\theta\tan^2\theta) & \sin^2\phi_2\tan^2\theta \\ \vdots & \vdots \\ \cos^2\phi_7 + \sin^2\phi_7(\cos^2\theta + \sin^2\theta\tan^2\theta) & \sin^2\phi_7\tan^2\theta \end{pmatrix}. \quad (31)$$

For practical use, the correction factor $\gamma$ is necessary for $\cos^2\phi_j$ ($j = 1, 2, \ldots, 7$) to account for the difference between $S_{\text{blank}}^{\text{s}}$ and $S_{\text{blank}}^{\text{p}}$ in the FT-IR spectrometer, which finally gives $\mathbf{R}$ in eq. 17 for IR-MAIRS.[32] For reference, the weighting factor matrix for pMAIRS can be also obtained by adopting $\phi = 90°$ for p-polarization in eq. 31:

$$\mathbf{R}_{\text{pMAIRS}} = \begin{pmatrix} \cos^2\theta_1 + \sin^2\theta_1\tan^2\theta_1 & \tan^2\theta_1 \\ \cos^2\theta_2 + \sin^2\theta_2\tan^2\theta_2 & \tan^2\theta_2 \\ \vdots & \vdots \\ \cos^2\theta_8 + \sin^2\theta_8\tan^2\theta_8 & \tan^2\theta_8 \end{pmatrix}. \quad (32)$$

**2.4. Analytical expressions for $A_{\text{IP}}$ and $A_{\text{OP}}$**

As shown above, the weighting factor matrix, $\mathbf{R}$, is heuristically deduced mainly through considering geometry rather than electromagnetism. As the linear combination part of a regression equation elucidates only a portion of the measured spectra, the $\mathbf{R}$ matrix cannot directly be deduced from Maxwell's equations. Instead, although it requires heavy mathematical calculations, the $\mathbf{A}_{\text{IP}}$ and $\mathbf{A}_{\text{OP}}$ spectra obtained by a given $\mathbf{R}$ matrix can be theoretically analyzed inductively with Maxwell's equations (classical electromagnetic theory), thus allowing evaluation of the validity of the $\mathbf{R}$ matrix for IR-MAIRS.[19,32]

According to Itoh et al.[19] and Shioya et al.[32], the $\boldsymbol{A}_{\text{IP}}$ and $\boldsymbol{A}_{\text{OP}}$ spectra obtained by the $\boldsymbol{R}$ matrix in eqs. 15 and 16 are expressed by the following linear combination of $f(\varepsilon)_{\text{TO}}$ and $f(\varepsilon)_{\text{LO}}$:

$$A_{\text{IP}} = \frac{8\pi d}{\lambda \ln 10}\left[h_{xy}^{\text{IP}} f(\varepsilon)_{\text{TO}} + h_{z}^{\text{IP}} f(\varepsilon)_{\text{LO}}\right], \qquad (33)$$

$$A_{\text{OP}} = \frac{8\pi d}{\lambda \ln 10}\left[h_{xy}^{\text{OP}} f(\varepsilon)_{\text{TO}} + h_{z}^{\text{OP}} f(\varepsilon)_{\text{LO}}\right], \qquad (34)$$

where $h_{xy}^{\text{IP}}$, $h_{z}^{\text{IP}}$, $h_{xy}^{\text{OP}}$, and $h_{z}^{\text{OP}}$ are weighting coefficients that depend on $\gamma$, $\theta$, $n_v$, and $n_s$; i.e., $h_{xy}^{\text{IP}}(\gamma, n_v, n_s, \theta)$, $h_{z}^{\text{IP}}(\gamma, n_v, n_s, \theta)$, $h_{xy}^{\text{OP}}(\gamma, n_v, n_s, \theta)$, and $h_{z}^{\text{OP}}(\gamma, n_v, n_s, \theta)$. In other words, they are all free from the optical parameters of a thin sample. As $h_{xy}^{\text{IP}}$, $h_{z}^{\text{IP}}$, $h_{xy}^{\text{OP}}$, and $h_{z}^{\text{OP}}$ have complicated forms, we refer interested readers to the original papers for their exact forms.[19,32] Nevertheless, their values can be calculated, and Fig. 12 shows the calculation results with respect to $\theta$ adopting $n_v = 1$ for vacuum conditions and $n_s = 3.41$ for a Si substrate.[32] Note that $\gamma$ can be separately corrected from experimental measurements as described in subsection 2.2. Figure 12 shows that the values for $h_{xy}^{\text{IP}}$ and $h_{z}^{\text{IP}}$ in $A_{\text{IP}}$ and for $h_{xy}^{\text{OP}}$, and $h_{z}^{\text{OP}}$ in $A_{\text{OP}}$ vary greatly with $\theta$, which means in principle that the $\boldsymbol{R}$ matrix in eq. 17 cannot solely decompose the measured seven oblique incidence spectra into two pure IP ($f(\varepsilon)_{\text{TO}}$) and OP ($f(\varepsilon)_{\text{LO}}$) spectra at all incident angles ($1° \leq \theta \leq 89°$) due to the undesirable contributions of $h_{z}^{\text{IP}} f(\varepsilon)_{\text{LO}}$ for $A_{\text{IP}}$ (eq. 33) and $h_{xy}^{\text{OP}} f(\varepsilon)_{\text{TO}}$ for $A_{\text{OP}}$ (eq. 34). However, Fig. 12 also shows that there is an optimal incidence angle ($\theta = 45°$ for Si) for IR-MAIRS that makes both the coefficients of $h_{z}^{\text{IP}}$ in $A_{\text{IP}}$ and $h_{xy}^{\text{OP}}$ in $A_{\text{OP}}$ become nearly zero; i.e., $h_{z}^{\text{IP}} \approx 0$ and $h_{xy}^{\text{OP}} \approx 0$.[30,32]

$$A_{\text{IP}} \approx \frac{8\pi d}{\lambda \ln 10} h_{xy}^{\text{IP}} f(\varepsilon)_{\text{TO}} = \frac{8\pi d}{\lambda \ln 10} h_{xy}^{\text{IP}} (2n_{xy}k_{xy}) \qquad (\theta = 45° \text{ for Si}) \qquad (35)$$

$$A_{\text{OP}} \approx \frac{8\pi d}{\lambda \ln 10} h_z^{\text{OP}} f(\varepsilon)_{\text{LO}} = \frac{8\pi d}{\lambda \ln 10} h_z^{\text{OP}} \frac{2n_z k_z}{(n_z^2 + k_z^2)^2} \qquad (\theta = 45° \text{ for Si}) \qquad (36)$$

Eqs. 35 and 36 show that $A_{\text{IP}}$ and $A_{\text{OP}}$ can be regarded as being proportional to $f(\varepsilon)_{\text{TO}}$ and $f(\varepsilon)_{\text{LO}}$, respectively, at the optimal incident angle of $\theta = 45°$ for a Si substrate. In other words, $f(\varepsilon)_{\text{TO}}$ and $f(\varepsilon)_{\text{LO}}$ can be simultaneously obtained from a single sample using IR-MAIRS. This unique characteristic of IR-MAIRS has significant advantages over conventional measurement techniques such as normal-incidence transmission and reflection–absorption using a metallic substrate. To represents this, the next section examines molecular orientation analysis as an example.

## 2.5. Molecular orientation analysis by IR-MAIRS

Using IR-MAIRS, quantitative molecular orientation analysis is easily performed for a thin sample. When the two approximations in eqs. 37 and 38 are satisfied, $A_{\text{IP}}$ and $A_{\text{OP}}$ in eqs. 35 and 36 are simplified to eqs. 39 and 40, respectively:

$$\frac{k^2}{n^2} \ll 1, (37)$$

$$n_{xy} = n_z = n, (38)$$

$$A_{\text{IP}} = \frac{8\pi d}{\lambda \ln 10} h_{xy}^{\text{IP}} (2n_{xy}k_{xy}) \approx \frac{8\pi d}{\lambda \ln 10} h_{xy}^{\text{IP}} (2nk_{xy}), (39)$$

$$A_{\text{OP}} = \frac{8\pi d}{\lambda \ln 10} h_z^{\text{OP}} \frac{2n_z k_z}{(n_z^2 + k_z^2)^2} \approx \frac{8\pi d}{\lambda \ln 10} \frac{h_z^{\text{OP}}}{n^4} (2nk_z). \quad (40)$$

The first approximation (eq. 37) indicates that the thin sample's extinction coefficient is small compared with the refractive index. The second approximation (eq. 38) indicates that the thin sample has no optical anisotropy and no anomalous dispersion, which is consistent with the first approximation.

Eqs. 39 and 40 straightforwardly indicate that the IP and OP spectra are proportional to $k_{xy}$ and $k_z$, respectively. Therefore, $k_{xy}$ and $k_z$ are derived as

$$k_{xy} = \frac{A_{\text{IP}} \lambda \ln 10}{16\pi dn h_{xy}^{\text{IP}}}, \quad (41)$$

$$k_z = \frac{A_{\text{OP}} n^4 \lambda \ln 10}{16\pi dn h_z^{\text{OP}}}; \quad (42)$$

their values for a thin sample reflect the orientation of a transition moment.[21,23] When a thin sample has only the surface-perpendicular component of a transition moment (i.e., when the molecular vibration in the thin sample is perfectly perpendicular to the substrate surface), $k_{xy} = 0$, and hence $k_z = 3k$. Similarly, when the molecular vibration in a thin sample is perfectly parallel to the substrate surface, we have $k_z = 0$, and hence $k_{xy} = \frac{3}{2} k$ in a uniaxial system ($k_x = k_y = k_{xy}$). Thus, the orientation angle of $\alpha$ in an ellipsoidal representation is expressed as follows using the optically isotropic (bulk) $k$ in eq. 1 (see also Fig. 13):[16]

$$\frac{\left(\sqrt{k_{xy}}\right)^2}{\left(\sqrt{\frac{3}{2}}k\right)^2} + \frac{\left(\sqrt{k_z}\right)^2}{\left(\sqrt{3k}\right)^2} = \frac{2k_{xy} + k_z}{3k} = 1, (43)$$

$$\sqrt{k_{xy}} = \sqrt{\frac{3}{2}}k\sin\alpha, (44)$$

$$\sqrt{k_z} = \sqrt{3k}\cos\alpha. (45)$$

Hence, optically isotropic $k$ in eq. 1 is related to $k_{xy}$ and $k_z$ by eq. 46:

$$k = \frac{2k_{xy} + k_z}{3}. (46)$$

By calculating $\tan\alpha$, the orientation angle of $\alpha$ is expressed in terms of $k_{xy}$ and $k_z$ as

$$\tan\alpha = \frac{\sin\alpha}{\cos\alpha} = \frac{\sqrt{\frac{2k_{xy}}{3}}}{\sqrt{\frac{k_z}{3}}} = \sqrt{\frac{2k_{xy}}{k_z}}, (47)$$

$$\alpha = \tan^{-1}\sqrt{\frac{2k_{xy}}{k_z}}. (48)$$

Eq. 48 means that the orientation angle can be obtained when both $k_{xy}$ and $k_z$ are available. As shown in eqs. 41 and 42, IR-MAIRS has the unique advantage that $k_{xy}$ and $k_z$ can be obtained from $A_{\text{IP}}$ and $A_{\text{OP}}$, respectively. Therefore, the orientation angle of $\alpha$ can be determined from eq. 48 using $A_{\text{IP}}$ and $A_{\text{OP}}$:

$$\alpha = \tan^{-1}\sqrt{\frac{\frac{2A_{\text{IP}}\lambda \ln 10}{16\pi dn h_{xy}^{\text{IP}}}}{\frac{A_{\text{OP}}n^4\lambda \ln 10}{16\pi dn h_z^{\text{OP}}}}} = \tan^{-1}\sqrt{\frac{2h_z^{\text{OP}}A_{\text{IP}}}{n^4 h_{xy}^{\text{IP}}A_{\text{OP}}}} = \tan^{-1}\sqrt{\frac{2A_{\text{IP}}}{n^4 H A_{\text{OP}}}}, (49)$$

$$H = \frac{h_{xy}^{\text{IP}}}{h_z^{\text{OP}}}, (50)$$

where $H$ is defined as a constant specific to the substrate used in the IR-MAIRS measurements. This substrate-specific constant acts to correct the intensity ratio of the electric fields at the optical interface along the surface-parallel and surface-perpendicular directions.[32] For a Si substrate at $\theta = 45°$, the values for $h_{xy}^{\text{IP}}$ and $h_z^{\text{OP}}$ are calculated as $h_{xy}^{\text{IP}} = 0.384$, and $h_z^{\text{OP}} = 1.150$ (as shown in Fig. 12),[32] which leads to $H = 0.334$. Shioya et al. report values of $H$ for different substrates.[32]. Eq. 49 indicates that $\alpha$ can be determined by comparing the IP and OP band intensities on the absorbance scale ($A_{\text{IP}}/A_{\text{OP}}$) when the refractive index of the thin sample ($n$) is available in advance. Note that the obtained orientation angle $\alpha$ in a thin sample is an average over the area irradiated by polarized IR light (about 20 mm diameter).

As described above, the measurement and analysis procedures of IR-MAIRS are complicated to some extent compared with those for conventional normal-incidence transmission and reflection–absorption measurements. Commercial accessories and software have recently been released for automated data collection and processing of IR-MAIRS with a FT-IR spectrometer, thus greatly expediting IR-MAIRS for a thin sample.[24,63]

## 3. Measurement results for vapor-deposited amorphous water

Figure 14 shows typical single-beam spectra for the background and sample in measurements of amorphous water on a Si substrate at 90 K. Seven single-beam spectra were measured at $\theta = 45°$ with seven polarization angles ($\phi$) from s-polarization ($\phi = 0°$) to p-polarization ($\phi = 90°$) in 15° steps. The spectra in Fig. 14A and B correspond to the $j$th single-beam spectrum, $S_{\text{obs},j}$ ($j = 1, 2, …, 7$), at a polarization angle of $\phi_j$ ($j = 1, 2, …, 7$) in the background and sample measurements, respectively. Using these spectra, the IP and OP spectra are calculated, as shown in Fig. 15, for (A) the dangling-OH peak at 3696 cm$^{-1}$ assigned to three-coordinated H$_2$O molecules located at the ice surface (Fig. 6), (B) the OH stretching band from four-coordinated H$_2$O molecules in the bulk ice, and (C) the decoupled OD stretching band from the HDO molecules in the bulk ice. The IR absorbance spectra measured by normal-incidence transmission are also shown in the figure for reference. In addition, Fig. 15D, E, and F shows the seven IR absorbance spectra calculated by eq. 51 using the $j$th single-beam spectra in the background and sample measurements ($S^{\text{bg}}_{\text{obs},j}$ and $S^{\text{sam}}_{\text{obs},j}$, respectively) at each polarization angle of $\phi_j$ ($j = 1, 2, …, 7$) to help understand IR-MAIRS:

$$A_j = -\log_{10} \frac{S^{\text{sam}}_{\text{obs},j}}{S^{\text{bg}}_{\text{obs},j}}, (51)$$

where $j = 1, 2, …, 7$ corresponds to $\phi = 0°, 15°, …, 90°$, respectively.

Before discussing the IP and OP spectra measured by IR-MAIRS in detail, it should be emphasized that no three-coordinated dangling-OH peak at 3696 cm$^{-1}$ is observed at 90 K in normal-incidence transmission (Fig. 15A) or in the s-polarization measurement ($\theta = 45°$, $\phi = 0°$) (Fig. 15D). These results show that the three-coordinated dangling OH bonds do not have surface-parallel vibration. As the polarization angle changed from s-polarization ($\phi = 0°$) to p-polarization ($\phi = 90°$), the peak at 3696 cm$^{-1}$ gradually appeared (Fig. 15D), clearly indicating that these bonds have surface-perpendicular vibration. Applying CLS regression to these oblique-incidence transmission measurements allows IR-MAIRS to give both pure IP and OP vibration spectra, as shown in Fig. 15A. As expected, the three-coordinated dangling-OH peak at 3696 cm$^{-1}$ appears only in the OP spectrum, while no clear peak appears in the IP spectrum. We thus conclude that the three-coordinated dangling-OH bonds are strongly (almost perfectly) oriented perpendicular to the ice surface at 90 K, giving an orientation angle of $\alpha \approx 0°$ from the surface normal (Fig. 6). Note that vapor-deposited amorphous water at 80–100 K has values of $k$ that are approximately two orders of magnitude smaller than $n$ at 3750–3650 cm$^{-1}$ in the bulk,[18] indicating that the two approximations for molecular orientation in eqs. 37 and 38 are satisfied; i.e., $\frac{k^2}{n^2} \ll 1$ and $n_{xy} = n_z = n$.

In the OH stretching vibrational region (Fig. 15B and E), the IP spectrum, normal-incidence transmission spectrum, and s-polarization spectrum ($\theta = 45°$, $\phi = 0°$) show similar band shapes, because these three spectra are expressed by $f(\varepsilon)_{\text{TO}}$. Changing from s-polarization ($\phi = 0°$) to p-polarization ($\phi = 90°$) made a broad shoulder peak appear at high-wavenumbers around 3400 cm$^{-1}$ (Fig. 15E), indicating that this shoulder peak originated from $f(\varepsilon)_{\text{LO}}$. As IR-MAIRS can separately obtain $f(\varepsilon)_{\text{TO}}$ and $f(\varepsilon)_{\text{LO}}$ as IP and OP spectra, respectively, the shape of $f(\varepsilon)_{\text{LO}}$ in the OH stretching vibrational band can be clarified as the OP spectrum shown in Fig. 15B. We also confirmed that the band shapes of $f(\varepsilon)_{\text{TO}}$ (IP spectrum) and $f(\varepsilon)_{\text{LO}}$ (OP spectrum) for amorphous water at 90 K obtained by IR-MAIRS are consistent with those at 4000–2800 cm$^{-1}$ calculated using reported values for the complex refractive index of amorphous water at 80 and 100 K (Fig. 16),[18] supporting the validity of IR-MAIRS. There is also good agreement between the IP and OP spectra of amorphous water at 10 K and the corresponding $f(\varepsilon)_{\text{TO}}$ and $f(\varepsilon)_{\text{LO}}$ calculated from the literature values, as shown in Figs. 4 and 5.[14,18] These results show that IR-MAIRS is potentially useful to study the TO–LO splitting of vapor-deposited amorphous water.[26,64]

Molecular orientation analysis requires a decoupled localized vibrational band whose transition moment direction reflects the molecular orientation. However, the OH stretching vibrations in H$_2$O ice and liquid are delocalized by intermolecular vibrational couplings, which complicates the interpretation of vibrational spectra for pure H$_2$O ice and liquid.[39–43,65,66] In addition, the TO–LO splitting in the OH stretching vibrational region observed in Figs. 15 and 16 indicates the dispersion of *n* induced by the large *k* values (i.e., the strong IR absorption). This means that the two approximations of $\frac{k^2}{n^2} \ll 1$ and $n_{xy} = n_z = n$ (eqs. 37 and 38) for molecular orientation no longer hold for the OH stretching band. Therefore, it is inappropriate to attempt molecular orientation analysis based on the OH stretching vibration band.

To obtain information about molecular orientation in the bulk of amorphous water, we focus on the decoupled-OD stretching band of dilute HDO (Fig. 15C and F).[39–43,65,66] This band is spectrally isolated and off-resonance with the major OH stretching band, and its concentration (3.5 mol% HDO) is sufficiently low for intermolecular vibrational coupling to be ignored. It has been widely used to investigate the bulk ice structure, because its peak wavenumber reflects the local lattice structure (oxygen–oxygen distance).[25,44–57] For example, crystalline ice I at 90 K shows a sharp single peak at 2417 cm$^{-1}$ in the decoupled OD stretching vibrational region.[25,44,51] The present normal-incidence transmission spectrum shows a broad peak at 2433 cm$^{-1}$ (Fig. 15C), confirming the formation of amorphous water at 90 K in this study. Figure 15F shows no clear difference among the absorbance spectra obtained between s-polarization ($\phi = 0°$) and p-polarization ($\phi = 90°$), similar to the IP and OP spectra from IR-MAIRS in Fig. 15C. These results are in sharp contrast to those for the OH stretching band (Fig. 15B and E). Repeated experiments for amorphous water samples at 90 K formed by vapor deposition for 32, 64, and 128 min at $2.2 \times 10^{-6}$ Pa show TO–LO splitting of $4 \pm 1$ cm$^{-1}$, considering the peak wavenumbers for the IP (2433.5 ± 0.5 cm$^{-1}$) and OP (2437.5 ± 0.5 cm$^{-1}$) spectra (see also Fig. 17B). This TO–LO splitting is significantly smaller than that for the OH stretching band, because the IP and OP spectra have peaks at around 3229 and 3381 cm$^{-1}$, respectively, which corresponds to TO–LO splitting of roughly 150 cm$^{-1}$ (Fig. 15B). Hence, the low concentration of HDO gives a weak extinction coefficient for the decoupled OD stretching vibrational band, and the dispersion of $n$ is much smaller than that for the OH stretching band. Therefore, the decoupled OD stretching band satisfies the two approximations of $\frac{k^2}{n^2} \ll 1$ and $n_{xy} = n_z = n$ (eqs. 37 and 38) necessary for molecular orientation analysis by IR-MAIRS. The following section discusses the temperature dependence of the orientations of dangling OH bonds on the ice surface and decoupled OD stretching in the bulk ice (Fig. 17).

## 4. Temperature dependence of dangling-OH and decoupled-OD bands

Vapor-deposited amorphous water has physical properties distinct from those of other crystalline water ices:[5] e.g., lower thermal conductivity,[67–69] larger surface area, and higher porosity.[70–75] Its porosity depends on several preparation parameters, including substrate temperature, deposition rate, method (background or molecular-beam deposition), and the incidence angle of the incoming water molecules.[70–73] For example, Dohnálek et al. reported that background deposition at 100 K results in nonporous amorphous water with a density of about 0.79 g cm$^{-3}$, whereas highly porous amorphous water with a density of about 0.62 g cm$^{-3}$ is formed at 22 K.[73] Stevenson et al. investigated the amount of $N_2$ adsorbed on amorphous water grown by background deposition, noting its increasing porosity with decreasing deposition temperature $\leq 77$ K.[70] They reported an apparent surface area of 2700 m$^2$ g$^{-1}$ at 22 K versus 640 m$^2$ g$^{-1}$ at 77 K.[70] Amorphous water prepared above about 90 K is reported to have almost identical $N_2$ adsorption to crystalline ice prepared at 140–145 K.[70,72]

Microscopic observation of nanoscale pores in vapor-deposited amorphous water remains challenging, especially at low temperatures around 10 K.[76–80] Instead, Garrod and co-workers recently used kinetic Monte Carlo simulations with an isotropic spherical molecules and the Lennard-Jones potential to investigate the porosity of vapor-deposited amorphous water.[74,81–83] For example, Fig. 18 shows side-views of the simulated ices produced at 10–120 K by background deposition with deposition rates of $10^1$, $10^{13}$, and $10^{16}$ molecules cm$^{-2}$ s$^{-1}$. The structure of the ice is clearly different at each temperature. At all deposition rates, the simulated ices have the highest porosity and large-scale irregular structures at 10 K, and they become smoother and less porous with increasing temperature. This trend qualitatively reproduces experimental observations of amorphous water surfaces.[70–74] These previous studies motivated us to use IR-MAIRS to study the dependence on ice temperature of the IP and OP spectra of dangling-OH and decoupled-OD stretching vibrations.

Figure 17 shows the IP and OP spectra of dangling-OH and decoupled-OD peaks in nonporous amorphous water at 90 K (A and B) and porous amorphous water at 10 K (C and D) with respect to gas exposure time. Similar to Fig. 15A, the OP spectra in Fig. 17A each show a single peak at 3696 cm$^{-1}$ for three-coordinated molecules at 90 K; however, the corresponding IP spectra do not show this peak. Furthermore, the intensity of the three-coordinated dangling-OH peak in the OP spectra remained constant at each exposure time for 32–128 min (Fig. 17A). This result suggests that three-coordinated dangling OH bonds in nonporous amorphous water at 90 K are located dominantly at the top ice surface and oriented perpendicular to it (Fig. 19). In contrast, the decoupled-OD peaks appear both in the IP and OP spectra at 90 K (Fig. 17B). The peak intensities clearly increase as the exposure time increases, because they reflect the properties of bulk ice. The average orientation angle, $\alpha$ (relative to the surface normal), of the decoupled-OD stretching vibration was calculated by eq. 49 using the peak areas in the IP and OP spectra at 2540–2380 cm$^{-1}$ and a reported value for the refractive index of amorphous water at 90 K ($n = 1.26$).[84] For exposure times of 32, 64, and 128 min at 90 K, $\alpha$ was calculated to be 58° ± 1°, 58°, and 58°, respectively. The error bars of $\alpha$ for the 64 and 128 min experiments are within ±1°. When the IP and OP spectra exhibit identical band intensities ($A_{IP} = n^4 HA_{OP}$), a perfectly random orientation angle $\alpha = 54.7°$ is obtained.[27] As the calculated angles are close to 54.7°, the present IR-MAIRS measurements show that the HDO molecules have an almost random orientation in the bulk ice at 90 K.

For porous amorphous water prepared at 10 K (Fig. 17C and D), both the IP and OP spectra of dangling OH bonds include two peaks at 3696 and 3720 cm$^{-1}$ (Fig. 17C). The latter is assigned to dangling OH bonds in two-coordinated (one H-donor and one H-acceptor) molecules located at the ice surface (Fig. 6).[35,36] The band intensities of both peaks increased as the ice thickness increased during gas deposition for 32–128 min. This behavior is clearly different from the three-coordinated dangling-OH bonds at 90 K (Fig. 17A). Figure 17C indicates that water vapor deposition at 10 K results in the formation of porous amorphous water with a large internal surface area (Fig. 19), in good agreement with previous experimental works[71–74,85] and with recent kinetic Monte Carlo simulations.[74,82]

Although overlap of the two dangling OH peaks hinders accurate calculation of the orientation angle $\alpha$, the similar band shapes and intensities for the IP and OP spectra at 10 K for each exposure time (32–128 min) indicate the isotropic nature (random orientation) of the dangling OH bonds in the internal pore surface (Fig. 17C). The intensity ratio of the decoupled-OD stretching band between the IP and OP spectra also reveals $\alpha = 56°$ from the surface normal irrespective of the exposure time (32–128 min), indicating almost perfectly random orientation (Fig. 17D). The error bars of $\alpha$ in these experiments are within ±1°. The broad peaks around 2450 cm$^{-1}$ for the IP and OP spectra of the decoupled-OD stretching band at 10 K suggest a more-disordered and larger oxygen–oxygen distance on average for amorphous water at 10 K than at 90 K (Fig. 14C and D).[49,50,52] Slightly larger $\alpha$ values at 90 K (58°) than at 10 K (56°) may imply the existence of a small molecular orientation anisotropy for amorphous water at 90 K, whereas more-quantitative discussion would require further study of, for example, the validity of using literature values for the refractive index of the present amorphous water samples.[84] This is a potential source of uncertainty in the molecular orientation analysis by IR-MAIRS (eq. 49). A more detailed investigation of the molecular orientation of vapor-deposited amorphous water in the bulk using the decoupled-OD stretching band from IR-MAIRS is currently underway.

## 5. Absorption cross-section of dangling OH bonds

Dangling OH bonds have been intensely studied to understand the surface properties of interstellar ice and its catalytic role in the formation of molecules in space, because they reflect the disorderliness of the ice surface structure as well as the ice's porosity, and also serve as catalytic sites for adsorption and chemical reactions.[5,70,84,86,87] Despite much active research, the absorption cross-section, a fundamental optical constant, remains unknown for dangling OH bonds. This hampers the quantification of dangling OH bonds from the IR spectra of ice, ultimately preventing a molecular-level understanding of the ice surface.

We first examine the difficulty in measuring the absorption cross-section of dangling OH bonds. The Beer–Lambert law (eq. 1) is described as follows with respect to absorption cross-section $\sigma$ (cm$^2$molecule$^{-1}$):

$$A = -\log_{10}\frac{S^s}{S^b} = \frac{4\pi dk}{\lambda \ln 10} = \frac{\sigma N}{\ln 10}, \quad (52)$$

where $N$ (molecule cm$^{-2}$) represents the column density of molecules in a sample. Hence, the absorption cross section $\sigma$ can be written as

$$\sigma = \frac{4\pi dk}{N\lambda}. \quad (53)$$

This shows that both the optically isotropic $k$ and column density $N$ of the sample are required to obtain $\sigma$. As optically isotropic $k$ is the extinction coefficient for a bulk sample, it is intrinsically difficult to define it for dangling OH bonds, which locate only on the surface and not in the bulk. Nevertheless, considering that optically isotropic $k$ is related to $k_{xy}$ and $k_z$ in eq. 46, a value for $k$ for dangling OH bonds can be obtained from quantitative measurement of both their IP ($k_{xy}$) and OP ($k_z$) vibration spectra. As described in the Introduction, however, $A_{\text{thin}}^{\theta=0°}$ for a normal-incidence transmission measurement of vapor-deposited amorphous water (a thin sample) only includes IP vibration given by $k_{xy}$ (eqs. 4–6). Likewise, $A_p^{\text{RA}}$ for reflection–absorption measurements includes only the OP vibration given by $k_z$ (eqs. 7 and 8). Moreover, $A_{\text{thin}}^{\theta=0°}$ (eqs. 4–6) and $A_p^{\text{RA}}$ (eqs. 7 and 8) are influenced by $n_v$, $n_s$, and $n_{xy}$, and by $n_v$, $\theta$, and $n_z$, respectively, indicating that they have different ordinate scales. These facts make quantitative measurement of $k_{xy}$ and $k_z$ challenging using current IR spectroscopy.

As shown below, we can purely derive $k_{xy}$ and $k_z$ for dangling OH bonds using IR-MAIRS by rearranging the analytical expressions for $A_{\mathrm{IP}}$ and $A_{\mathrm{OP}}$, which eventually leads to the optically isotropic $k$ and absolute value of $\sigma$. For porous amorphous water at 10 K, the two dangling OH peaks both overlap in the IP and OP spectra, which complicates measurement of $\sigma$ (Fig. 17C). In the following, we thus focus on non-porous amorphous water at 90 K, because it exhibits only the three-coordinated dangling-OH peaks (Fig. 17A). Itoh et al. and Shioya et al.[19,32] ascertained that $A_{\mathrm{IP}}$ in eq. 35 corresponds quantitatively to the absorbance in the s-polarized transmission measurement at the incidence angle of $\theta$ ($A_{\mathrm{thin_s}}^{\theta=45°}$) (eqs. 54 and 55), because $h_{xy}^{\mathrm{IP}}$ becomes numerically identical to $a'$:

$$A_{\mathrm{IP}} = \frac{8\pi d}{\lambda \ln 10} h_{xy}^{\mathrm{IP}}(2n_{xy}k_{xy}) \approx \frac{8\pi d a'}{\lambda \ln 10}(2n_{xy}k_{xy}) = A_{\mathrm{thin_s}}^{\theta=45°}, \quad (54)$$

$$h_{xy}^{\mathrm{IP}} \approx a' = \frac{1}{n_v \cos\theta + n_s \cos\theta_3} \\ + \left(\frac{n_v \cos\theta - n_s \cos\theta_3}{n_v \cos\theta + n_s \cos\theta_3}\right)^4 \left\{1 \\ - \left(\frac{n_v \cos\theta - n_s \cos\theta_3}{n_v \cos\theta + n_s \cos\theta_3}\right)^4\right\}^{-1} \left(\frac{2n_s \cos\theta_3}{n_s^2 - n_v^2}\right). \quad (55)$$

In fact, $h_{xy}^{IP} = a' = 0.3840$ when adopting $\theta = 45°$, $n_v = 1$, and $n_s = 3.41$ (Fig. 12). Thus, when absorbance correction is performed for $A_{\mathrm{IP}}$ in eq. 54 by calculating the intensity ratio between $a'$ in eq. 55 and $a$ in eq. 5, $\frac{a}{a'}A_{\mathrm{IP}}\left(A_{\mathrm{IP}}^{\theta=0°}\right)$ coincides with the absorbance in the normal-incidence transmission measurement ($A_{\mathrm{thin}}^{\theta=0°}$ in eq. 4):

$$A_{\mathrm{IP}}^{\theta=0°} = \frac{a}{a'}A_{\mathrm{IP}} = \frac{8\pi d a}{\lambda \ln 10}(2n_{xy}k_{xy}) = A_{\mathrm{thin}}^{\theta=0°}. \quad (56)$$

As the value for $a$ in eq. 5 is calculated to be 0.2896 for a Si substrate in a vacuum ($n_v = 1$ and $n_s = 3.41$), the intensity ratio of $a/a' = 0.2896/0.3840 = 1/1.326 = 0.754$. We confirm eq. 56 by comparing $A_{IP}^{\theta=0°}$ and $A_{thin}^{\theta=0°}$ for the decoupled-OD stretching band at 90 K (Fig. 15C), because the three-coordinated dangling-OH bonds only have the OP vibration at 90 K (Figs. 15A and 17 A). Table 2 summarizes the band shape and intensity for the decoupled-OD stretching vibrational band in the IP spectrum ($A_{IP}^{\theta=0°}$) and normal-incidence spectrum ($A_{thin}^{\theta=0°}$) shown in Fig. 15C. The IP spectrum and normal-incidence spectrum for amorphous water at 90 K have almost identical decoupled-OD stretching bands in terms of peak wavenumber, height, width, and area within an error of 1%. This validates our experimental IR-MAIRS measurements and analysis.

Following the same absorbance correction for $A_{OP}$ in eq. 36, $\frac{a}{a'} A_{OP}(A_{OP}^{\theta=0°})$ is obtained as follows:

$$A_{OP}^{\theta=0°} = \frac{a}{a'} A_{OP} = \frac{a}{a'} \frac{8\pi d}{\lambda \ln 10} h_z^{OP} \frac{2 n_z k_z}{(n_z^2 + k_z^2)^2}. \quad (57)$$

As $H = \frac{h_{xy}^{IP}}{h_z^{OP}}$ (eq. 50) and $h_{xy}^{IP} \approx a'$ (eq. 55), $h_z^{OP} = \frac{h_{xy}^{IP}}{H} \approx \frac{a'}{H}$ and the term $\frac{h_z^{OP}}{a'}$ in $A_{OP}^{\theta=0°}$ can be replaced with $\frac{1}{H}$ as follows:

$$A_{OP}^{\theta=0°} \approx \frac{a}{a'} \frac{8\pi d}{\lambda \ln 10} \frac{a'}{H} \frac{2 n_z k_z}{(n_z^2 + k_z^2)^2} = \frac{8\pi d a}{\lambda H \ln 10} \frac{2 n_z k_z}{(n_z^2 + k_z^2)^2}. \quad (58)$$

Calculating eqs. 56 and 58 is practically convenient, because the coefficient $a$ only depends on $n_v$ and $n_s$ (eq. 3). In addition, under the approximations of $\frac{k^2}{n^2} \ll 1$ and $n_{xy} = n_z = n$ (eqs. 37 and 38), $A_{IP}^{\theta=0°}$ in eq. 56 and $A_{OP}^{\theta=0°}$ in eq. 58 are further simplified, and eventually become proportional to $k_{xy}$ and $k_z$, respectively (see also subsection 2.5):

$$A_{IP}^{\theta=0°} \approx \frac{8\pi da}{\lambda \ln 10}(2nk_{xy}), (59)$$

$$A_{OP}^{\theta=0°} = \frac{8\pi da}{\lambda H \ln 10}\frac{2n_z k_z}{(n_z^2 + k_z^2)^2} \approx \frac{8\pi da}{\lambda H \ln 10}\frac{2nk_z}{n^4} = \frac{8\pi da}{n^4 H \lambda \ln 10}(2nk_z). (60)$$

These equations allow pure $k_{xy}$ and $k_z$ to be expressed with a common denominator of $16\pi dna$:

$$k_{xy} = \frac{A_{IP}^{\theta=0°} \lambda \ln 10}{16\pi dna}, (61)$$

$$k_z = \frac{n^4 H A_{OP}^{\theta=0°} \lambda \ln 10}{16\pi dna}. (62)$$

As optically isotropic $k$ is related to $k_{xy}$ and $k_z$ in eq. 46, it can also be simply expressed in terms of $A_{IP}^{\theta=0°}$ and $A_{OP}^{\theta=0°}$:[27]

$$k = \frac{2k_{xy} + k_z}{3} = \frac{2}{3}\left(\frac{A_{IP}^{\theta=0°}\lambda \ln 10}{16\pi dna}\right) + \frac{1}{3}\left(\frac{n^4 H A_{OP}^{\theta=0°}\lambda \ln 10}{16\pi dna}\right)$$

$$= \frac{\lambda \ln 10}{16\pi dna}\left(\frac{2A_{IP}^{\theta=0°} + n^4 H A_{OP}^{\theta=0°}}{3}\right). (63)$$

Therefore, $\sigma$ (eq. 53) can be experimentally measured by IR-MAIRS, if $N$ is known:

$$\sigma = \frac{4\pi dk}{N\lambda} = \frac{4\pi d}{N\lambda}\frac{\lambda \ln 10}{16\pi dna}\left(\frac{2A_{IP}^{\theta=0°} + n^4 H A_{OP}^{\theta=0°}}{3}\right) = \frac{\ln 10}{4naN}\left(\frac{2A_{IP}^{\theta=0°} + n^4 H A_{OP}^{\theta=0°}}{3}\right). (64)$$

The above discussion shows that quantifying the column density $N$ of dangling-OH bonds is crucial to obtaining the absorption cross-section. For reference, the OMNIC software for IR-MAIRS analysis gives $A_{IP}^{\theta=0°}$ and $n^4 H A_{OP}^{\theta=0°}$ as the ordinate scales for the resultant IP and OP spectra, respectively.[63]

Nagasawa et al.[34] estimated the column density of three-coordinated dangling-OH bonds at 90 K using CH$_3$OH deposition, because its hydrogen bonding with dangling OH bonds quenches the dangling-OH peak at 3696 cm$^{-1}$.[88] Figure 20A and B shows the IP and OP spectra of amorphous water at 90 K before and after CH$_3$OH deposition in the ranges of (A) 3750–3660 cm$^{-1}$ for the dangling-OH peak and (B) 1060–1010 cm$^{-1}$ for the CO stretching band of CH$_3$OH. The pressure in the chamber was $2.2 \pm 0.2 \times 10^{-8}$ Pa during CH$_3$OH exposure, which corresponds to a flux of $7.8 \pm 0.9 \times 10^{10}$ molecules cm$^{-2}$ s$^{-1}$. The OP spectra following CH$_3$OH exposure show the three-coordinated dangling-OH peak at 3696 cm$^{-1}$ clearly decreasing after 2.5 min of deposition, and completely disappearing after 6.5 min (Fig. 20A). A new peak appeared at 1033 cm$^{-1}$ only in the OP spectra after CH$_3$OH deposition (Fig. 20B). Bahr et al. assigned the peak at 1034 cm$^{-1}$ to the CO stretching band of CH$_3$OH species directly interacting with the amorphous water surface through hydrogen bonding.[88] Considering the appearance of the CO stretching band of CH$_3$OH only in the OP spectra, the CO band has a strong orientation perpendicular to the amorphous water surface, probably through hydrogen-bond formation with the three-coordinated dangling OH bonds (Fig. 21).

Nagasawa et al. reported that CH$_3$OH deposition for $6.5 \pm 0.5$ min is required to fully quench the peak, which corresponds to an exposure of $3.1 \pm 0.4 \times 10^{13}$ molecules cm$^{-2}$.[34] They also calculated the column density of CH$_3$OH adsorbed on the amorphous water surface from the C–O stretching band at 1033 cm$^{-1}$ (Fig. 20B).[34] Eq. 64 shows that, when the band strength $\beta$ (the integrated absorption cross-section) for the C–O stretching band of CH$_3$OH is available, the column density $N$ of CH$_3$OH can be obtained from the IP and OP peak areas at $\tilde{v} = 1050$–$1020$ cm$^{-1}$:

$$\beta = \int \sigma(\tilde{v})d\tilde{v} = \frac{\ln 10}{4naN}\left(\frac{2}{3}\int A_{IP}^{\theta=0°}(\tilde{v})d\tilde{v} + \frac{n^4H}{3}\int A_{OP}^{\theta=0°}(\tilde{v})d\tilde{v}\right), \quad (65)$$

$$N = \frac{\ln 10}{4na\beta}\left(\frac{2}{3}\int A_{IP}^{\theta=0°}(\tilde{v})d\tilde{v} + \frac{n^4H}{3}\int A_{OP}^{\theta=0°}(\tilde{v})d\tilde{v}\right), \quad (66)$$

where $A_{IP}^{\theta=0°}(\tilde{v})$ and $A_{OP}^{\theta=0°}(\tilde{v})$ are the IP and OP absorbance at a given wavenumber ($\tilde{v}$), respectively. Luna et al. reported the value for $\beta$ in the CO stretching band of amorphous CH$_3$OH as $\beta = 1.59$–$1.69 \times 10^{-17}$ cm molecule$^{-1}$.[89] As CH$_3$OH is adsorbed on both sides of the Si substrate, the values for $\int A_{IP}^{\theta=0°}(\tilde{v})d\tilde{v}$ and $n^4H \int A_{OP}^{\theta=0°}(\tilde{v})d\tilde{v}$ for the CO stretching band are calculated as 0 cm$^{-1}$ and $1.1 \pm 0.1 \times 10^{-3}$ cm$^{-1}$, respectively, after $6.5 \pm 0.5$ min of CH$_3$OH deposition in repeated experiments. Hence, the column density of CH$_3$OH is calculated as $N = 3.5 \pm 0.4 \times 10^{13}$ molecules cm$^{-2}$, which is in good agreement with the amount of CH$_3$OH exposure ($3.1 \pm 0.4 \times 10^{13}$ molecules cm$^{-2}$). Note that as the coverage of CH$_3$OH (less than 0.1 monolayer) is much smaller than that of amorphous water (about 15 monolayers), the refractive index of amorphous water at 90 K (1.26) is used as a good approximation for $n$ in eq. 66.[33,34,84] As a single peak for the CO stretching band only appeared in the OP spectra at 1033 cm$^{-1}$ (Fig. 20B), all CH$_3$OH deposited on the amorphous water surface should interact with the three-coordinated dangling-OH bonds, leading to the strong orientation of CO bonds perpendicular to the surface (Fig. 21). Hence, we assume that the amount of $CH_3OH$ deposition required to fully quench the peak reflects the number density of three-coordinated dangling-OH bonds at 90 K, and we estimate the number density of three-coordinated dangling-OH bonds at 90 K to be $3.3 \pm 0.6 \times 10^{13}$ molecules $cm^{-2}$.

Seven independent measurements for amorphous water formed after 32 min of exposure at 90 K gave $A_{IP}^{\theta=0°} = 0$ [$\int A_{IP}^{\theta=0°}(\tilde{v})d\tilde{v} = 0$ $(cm^{-1})$] and $n^4 H A_{OP}^{\theta=0°} = 6.2 \pm 0.4 \times 10^{-5}$ at 3696 $cm^{-1}$ [$n^4 H \int A_{OP}^{\theta=0°}(\tilde{v})d\tilde{v} = 8.4 \pm 0.4 \times 10^{-4}$ $(cm^{-1})$], considering amorphous water formed on both sides of the Si substrate. The contribution of the surface HDO molecules to dangling OH bonds should be negligible considering its small concentration of 3.5 mol%. Therefore, $\sigma$ for the three-coordinated dangling-OH bonds is derived as $1.0 \pm 0.2 \times 10^{-18}$ $cm^2$ at 3696 $cm^{-1}$ from eq. 64 by adopting $N = 3.3 \pm 0.6 \times 10^{13}$ molecules $cm^{-2}$. The corresponding value for $\beta$ is derived as $1.4 \pm 0.3 \times 10^{-17}$ cm molecule$^{-1}$ at $\tilde{v} = 3710$–$3680$ $cm^{-1}$ from eq. 65. We emphasize that the three-coordinated dangling-OH peaks and the CO stretching bands at 90 K in the experiments described above should be invisible to normal-incidence transmission measurements, which observe only the IP vibration without any OP vibration contribution. The unique property of IR-MAIRS allows quantitative discussion of a strongly orientated species in a thin sample, such as dangling-OH bonds.

Table 3 summarizes band strength values for water. Our experimental values ($1.4 \pm 0.3 \times 10^{-17}$ cm molecule$^{-1}$) are consistent with a theoretical value recently calculated by Maté et al. ($1.2 \times 10^{-17}$ cm molecule$^{-1}$).[90] The three-coordinated dangling-OH bonds at 90 K show more than one order of magnitude smaller band strengths than those for the OH stretching bands of four-coordinated molecules in bulk amorphous and crystalline ices and liquid water.[18,34,91,92] This indicates that the lack of hydrogen bonding in the dangling OH bonds of three-coordinated molecules significantly decreases the band strength.[90,93,94] Table 3 also shows the band strengths reported by Ehrenfreund et al. for the anti-symmetric OH stretching vibration of $H_2O$ monomers confined in solid matrices of CO, $N_2$, and $O_2$,[95] which range from $3.3 \times 10^{-18}$ to $1.1 \times 10^{-17}$ cm molecule$^{-1}$ depending on the chemical components of the solid matrices.[90,96] These values are smaller than the band strength of the three-coordinated dangling-OH bonds ($1.4 \pm 0.3 \times 10^{-17}$ cm molecule$^{-1}$) measured by IR-MAIRS. We speculate that the dipole moment of a three-coordinated molecule at the amorphous water surface may be enhanced by surrounding water molecules in comparison with $H_2O$ monomers confined in solid matrices,[97,98] which leads to an increase in the band strength of the three-coordinated dangling-OH bonds. Although the dangling OH features have yet to be observed in the IR spectra of interstellar ices,[6] the James Webb Space Telescope will search with high sensitivity for dangling OH bonds in interstellar ices. The absorption cross-section reported here allows the quantification of dangling OH bonds and clarification of the surface structure and properties of interstellar ices and their laboratory analogues.

**Conclusions and perspectives**

This paper demonstrates that IR-MAIRS can provide new insights into the structure and properties of vapor-deposited amorphous water at the molecular level. As specific examples, the average molecular orientation of dangling OH bonds as well as their absorption cross-section are quantified using IR-MAIRS. $CH_3OH$ deposition experiments quantified the low column density (on the order of $10^{13}$ molecules cm$^{-2}$) of three-coordinated dangling OH bonds on the surface at 90 K. The information presented here should advance our quantitative understanding of the surface structure of interstellar ices and their laboratory analogs. FT-IR spectrometers are now widely used in most laboratories studying interstellar ices. Nevertheless, IR-MAIRS is yet to be commonly used in laboratory astrochemistry, and its potential has not previously been well explored. As a further study of dangling OH bonds by IR-MAIRS, measurements of $\sigma$ and band strength for the two-coordinated dangling-OH bonds are currently in progress to understand the effects of the hydrogen-bond coordination number. In addition to the peaks for dangling-OH and decoupled-OD, the OH stretching and libration bands in vapor-deposited amorphous water are also complex, but fascinating, topics. For example, the origin of TO–LO splitting in the OH stretching and libration bands remains poorly understood, and the extraction of physically meaningful information from this splitting is yet to be fully discussed (Figs. 5, 15, and 22). Because the libration band has been used to identify ice XI, the hydrogen atom-ordered form of ice Ih, IR-MAIRS study of the libration band can help astronomical IR observations to find ice XI in space.[99,100] The non-contact and non-destructive nature of IR-MAIRS will also facilitate the simultaneous and subsequent use of other analytical tools such as mass spectrometry. Combination studies would quantitatively reveal the kinetics and reaction dynamics of chemical reactions occurring in or on interstellar ice analogues, which would eventually improve our understanding of the chemical evolution of star- and planet-forming regions in space.

**Acknowledgements**

We thank Profs. Takeshi Hasegawa, Takafumi Shimoaka, and Nobutaka Shioya at the Institute for Chemical Research, Kyoto University, for useful suggestions regarding the measurements and analysis of IR-MAIRS. We also thank our colleagues at the Institute of Low Temperature Science, Hokkaido University, especially Profs. Naoki Watanabe and Akira Kouchi, for supporting the development of low-temperature, ultrahigh-vacuum IR-MAIRS. Some of the authors' studies described in this review were supported by JSPS Kakenhi grant numbers 18H01262, 19K22901, 21H01143, and 21H05421.

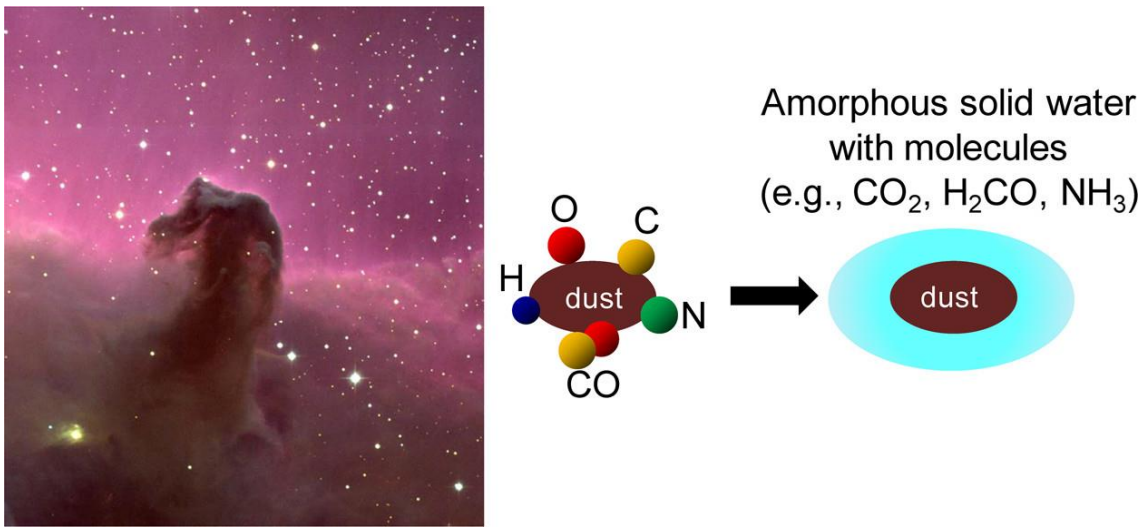

Fig. 1. (Left) Part of the Orion molecular cloud containing the Horsehead Nebula (also known as Barnard 33). The red glow behind the Horsehead is the emission nebula IC 434, which originates from hydrogen gas ionized by the bright star σ Orionis. The darkness of the Horsehead is caused mostly by dust grains, which are not penetrated by ultraviolet and visible photons from external stars. Bright spots at the base of the Horsehead are young stars in the process of forming. (Credit and copyright: N. A. Sharp/NOAO/AURA/NSF.) (Right) Schematic description of the morphological and chemical structure of dust grains. After the temperature and photon field decrease when the density of dust particles increases in dense molecular clouds, atoms (e.g., H, O, C, N) and molecules (e.g., CO) are deposited onto the dust surfaces. Cold-surface reactions proceed on the surfaces, and water-dominated ice is formed. Reprinted from ref.[5] with permission. Copyright 2013 American Chemical Society.

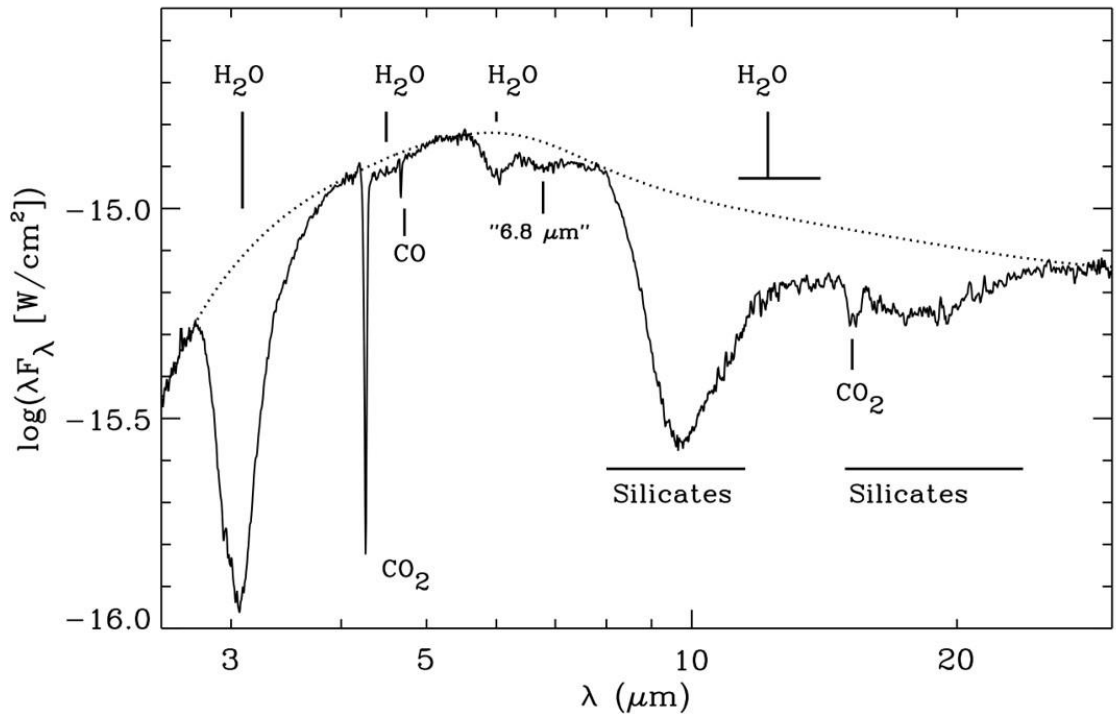

Fig. 2. Infrared spectrum obtained from observation of the low-mass protostar Elias 29 by the Infrared Space Observatory–Short Wavelength Spectrometer (ISO–SWS).[12] The protostar is located in the ρ Ophiuchi molecular cloud. The dotted line indicates a blackbody continuum. The vibrational absorption bands of various molecules are indicated. The "6.8 μm" band is an unknown band possibly originating from multiple components such as $NH_4^+$ and $CH_3OH$.[11] Credit: Boogert et al., A&A, 360, 683, 2000, reproduced with permission. Copyright ESO.

Table 1. Ice composition in a low-mass protostar (Elias 29) and comet (Hale–Bopp) expressed as a percentage of the $H_2O$ content[a]

| Molecule | Low-mass protostar (Elias 29)[b] (Column density [$10^{17}$ molecules cm$^{-2}$]) | Comet (Hale–Bopp)[c] |
|---|---|---|
| $H_2O$ | 100.0 (34) | 100 |
| $CO_2$ | 19.7 (6.7) | 20 |
| CO | 5.0 (1.7) | 20 |
| $CH_4$ | <1.5 (<0.5)[d] | 0.6 |
| $NH_3$ | <10.3 (<3.5)[d] | 0.6 |
| $CH_3OH$ | <4.4 (<1.5)[d] | 2 |
| $H_2CO$ | <1.8 (<0.6)[d] | 0.1–1.0 |
| HCOOH | <0.9 (<0.3)[d] | 0.05 |
| OCS | <0.04 (<0.015)[d] | |
| XCN[e] | <0.2 (<0.067)[d] | |

[a] Ref.[11] gives more details on the molecular abundances in interstellar ice mantles observed through IR absorption.

[b] Ref.[12]. Elias 29 is a low-mass protostar (i.e., having the mass of the Sun or lower) located in the ρ Ophiuchi molecular cloud.

[c] Ref.[13]

[d] Upper limits with 3σ standard deviations.

[e] C≡N stretch possibly due to OCN⁻.[101]

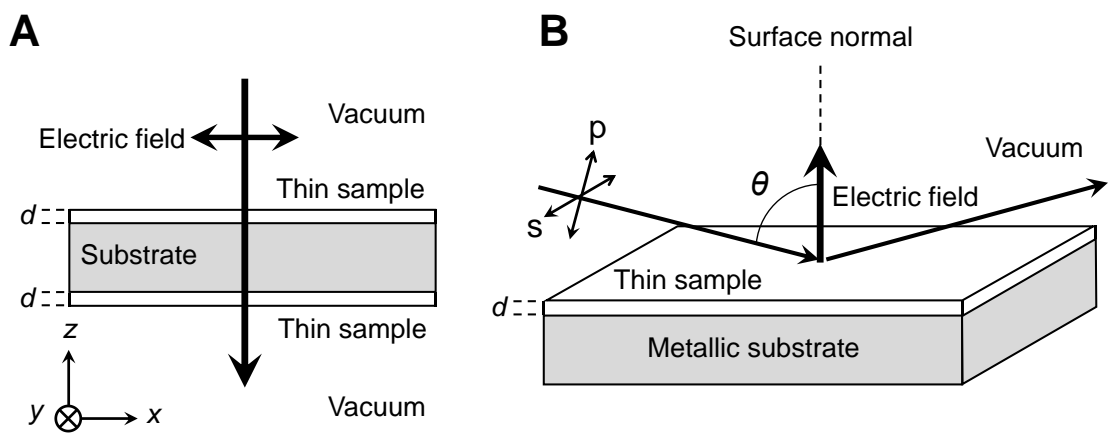

Fig. 3. Schematics showing commonly used IR measurements for a thin sample of thickness *d*. (a) Normal-incidence transmission measurement using an IR transparent substrate, and (b) reflection–absorption measurements using a metallic substrate. $\theta$ is the angle of incidence. ⊗ indicates the *y*-axis direction parallel to the substrate and perpendicular to the incident plane.

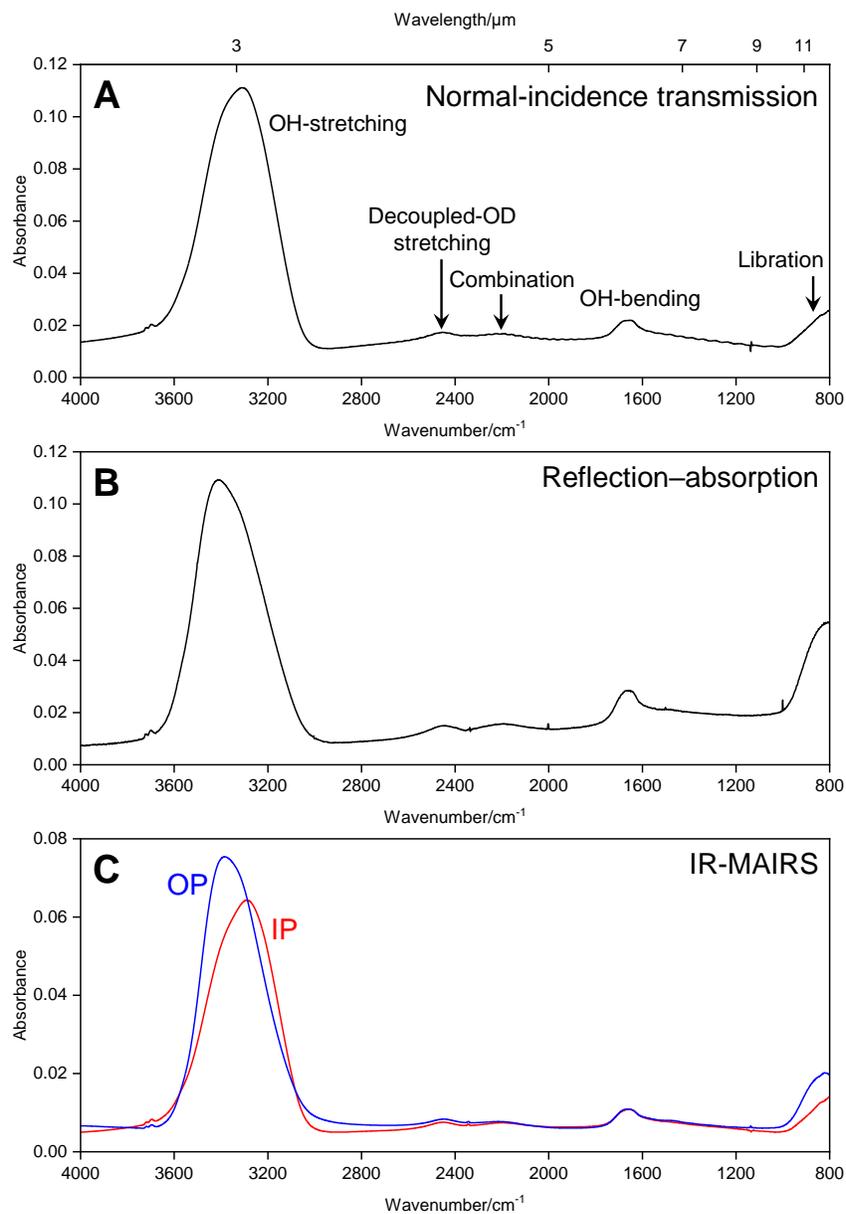

Fig. 4. Infrared spectra of vapor-deposited amorphous water at 4000–800 cm$^{-1}$. (A) Normal-incidence measurement on a Si substrate at 10 K. (B) Reflection–absorption measurement on an Al substrate at 11 K. Reproduced from Ref.[25] with permission from the Royal Society of Chemistry. (C) In-plane (IP) and out-of-plane (OP) spectra obtained by IR-MAIRS for vapor-deposited amorphous water on a Si substrate at 10 K prepared by 128 min exposure of water at $2.2 \times 10^{-6}$ Pa. The amorphous water samples in (A)–(C) were prepared using $H_2O$ with 3.5 mol% HDO.

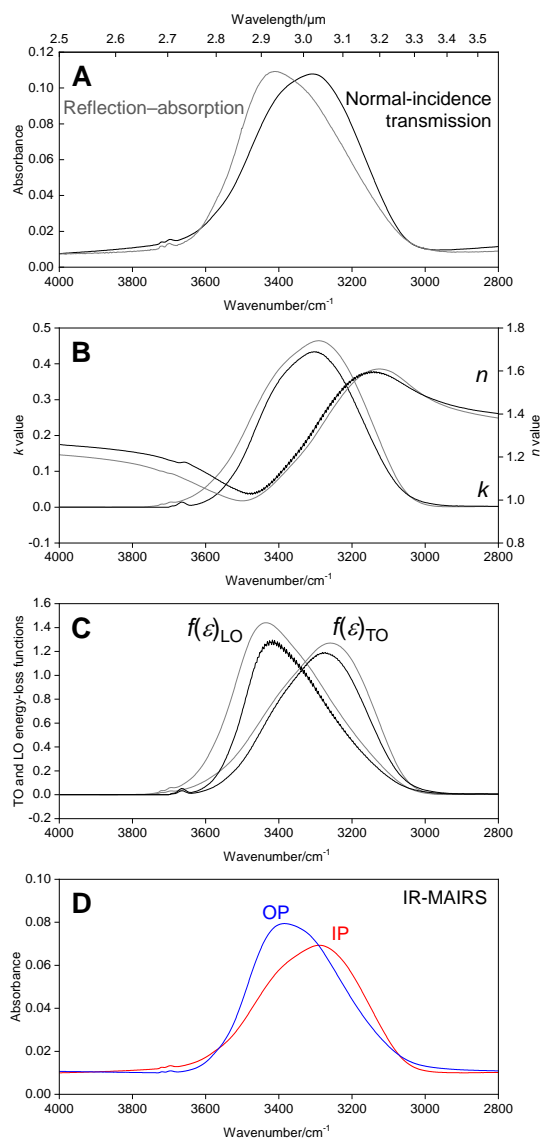

Fig. 5. Infrared spectra of vapor-deposited amorphous water in the OH-stretching vibrational region at 4000–2800 cm$^{-1}$. (A) Normal-incidence measurement at 10 K (black), and reflection–absorption measurement at 11 K (gray). The spectra are identical to those in Fig. 4A and B, respectively. Reproduced from Ref.[25] with permission from the Royal Society of Chemistry. (B) Literature values for the real ($n$) and imaginary ($k$) parts of the complex refractive index of amorphous water at 10 K (black, ref.[14]) and at 15 K (gray, ref.[18]). (C) Transverse optic (TO) energy-loss function ($f(\varepsilon)_{TO}$) and longitudinal optic (LO) energy-loss functions ($f(\varepsilon)_{LO}$) calculated with the $n$ and $k$ values at 10 K (black, ref.[14]) and at 15 K (gray, ref.[18]). The values for $f(\varepsilon)_{LO}$ are multiplied by three for clarity. (D) In-plane (IP) and out-of-plane (OP) spectra obtained by IR-MAIRS for vapor-deposited amorphous water at 10 K. The spectra are identical to those in Fig. 4C.

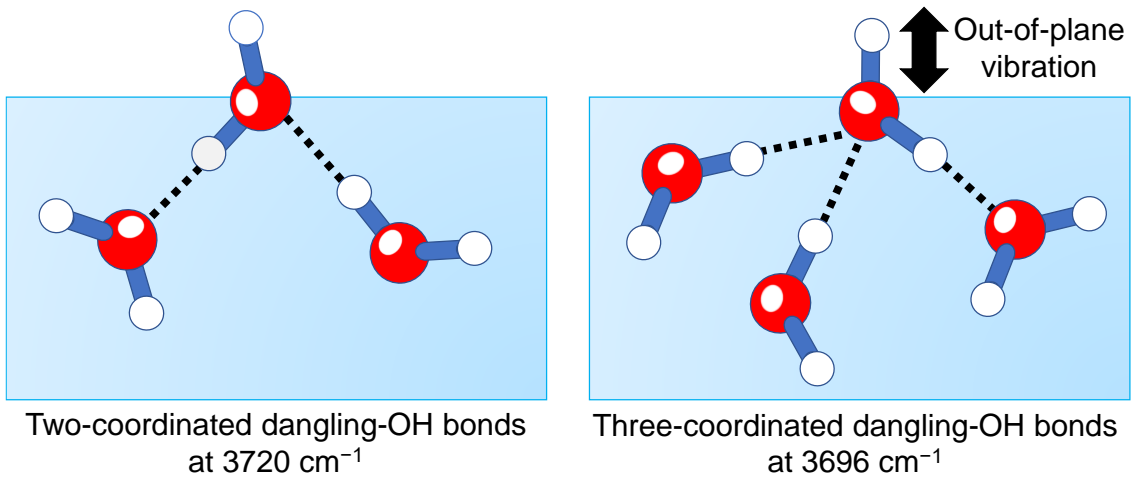

Fig. 6. Schematics of two- and three-coordinated dangling-OH bonds at the surface of water ice. IR-MAIRS reveals that the three-coordinated dangling-OH bonds have a strong out-of-plane vibration at 90 K (see also Figs. 15 and 17).

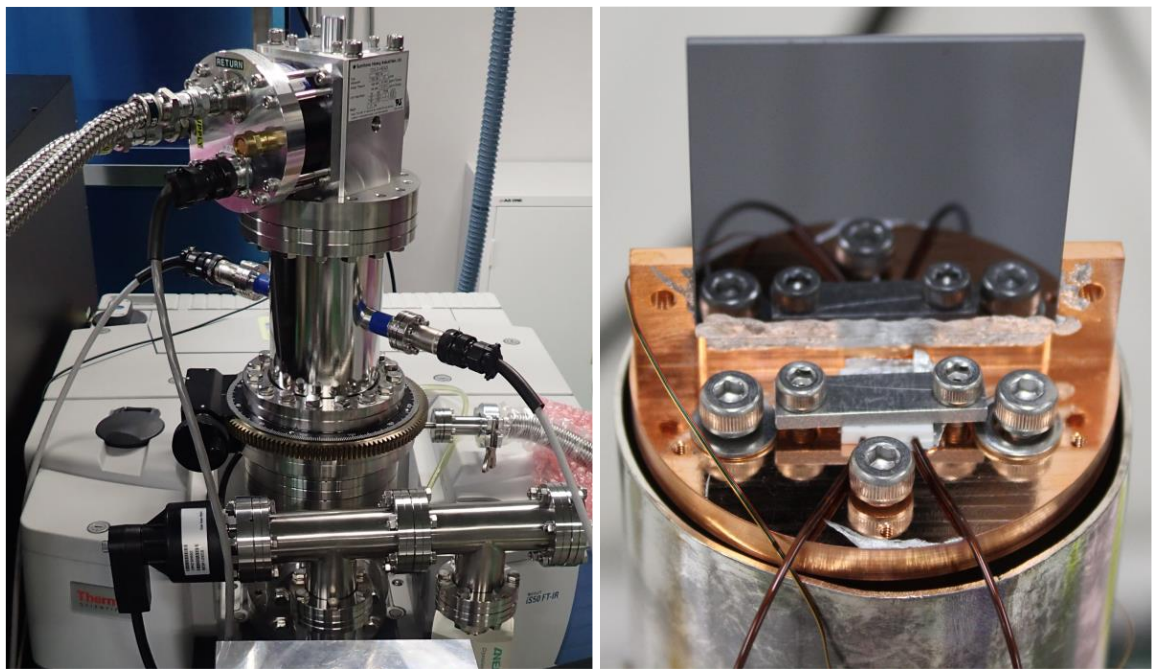

Fig. 7. Photographs of equipment for low-temperature, ultrahigh-vacuum IR-MAIRS. (left) Overall view. (right) Si substrate connected to a copper sample holder by ultrasonic-soldered indium. A ceramic heater (40 W) is located in front of the Si substrate. Reprinted from ref.[33] with permission. Copyright 2020 American Chemical Society.

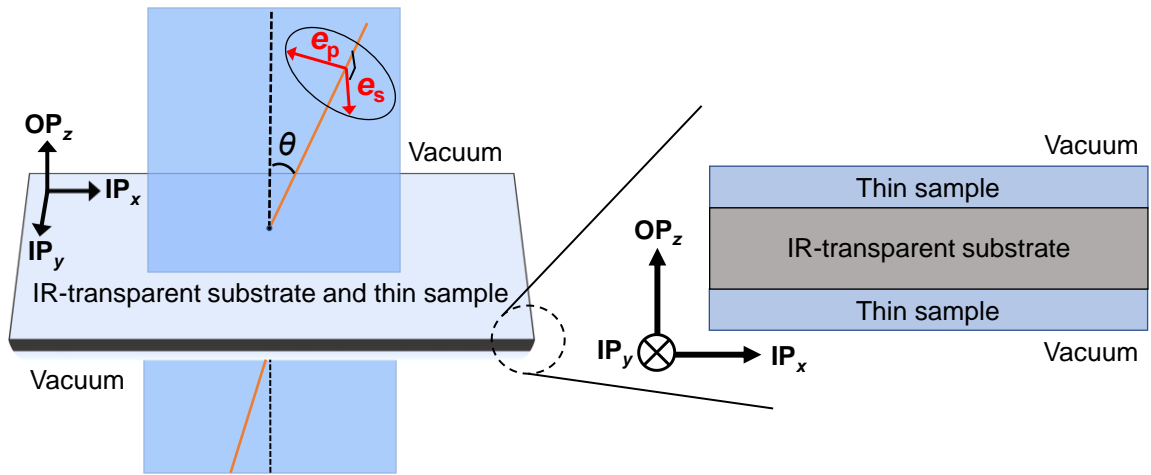

Fig. 8. Schematics of the IR-MAIRS system in a five-layer system (vacuum/thin sample/IR-transparent substrate/thin sample/vacuum). $e_s$ and $e_p$ represent the s-polarized and p-polarized basis vectors, respectively. $\theta$ is the angle of incidence. ⊗ indicates the direction of IP$_y$-axis parallel to the substrate and perpendicular to the incident plane.

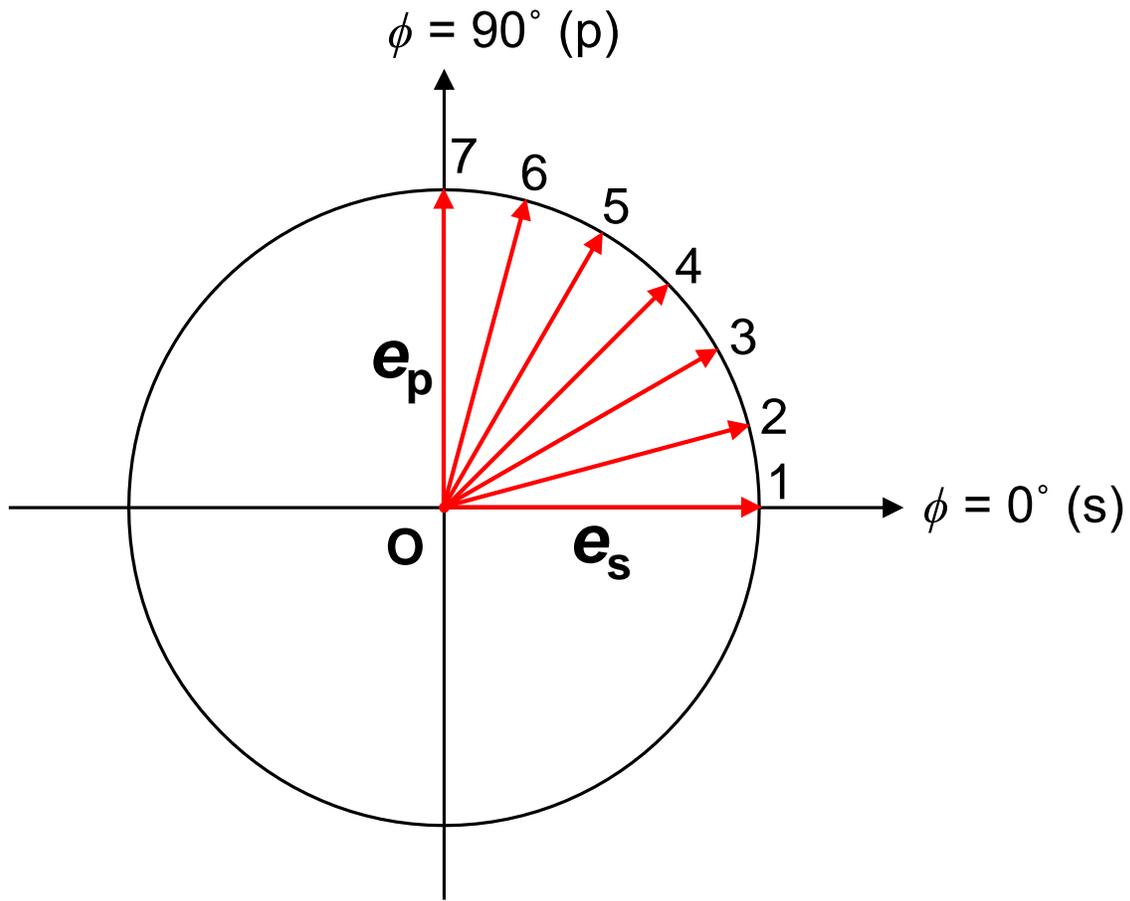

Fig. 9. Schematic of the polarization points chosen for IR-MAIRS measurements. The polarization angles of $\phi = 0°$ and $90°$ correspond to p- and s-polarization, respectively. $e_s$ and $e_p$ represent the s- and p-polarized basis vectors, respectively. In IR-MAIRS, oblique-incidence transmission measurements are performed at an incidence angle of $\theta = 45°$ at seven polarization angles from $\phi = 0°$ to $90°$ in $15°$ steps.

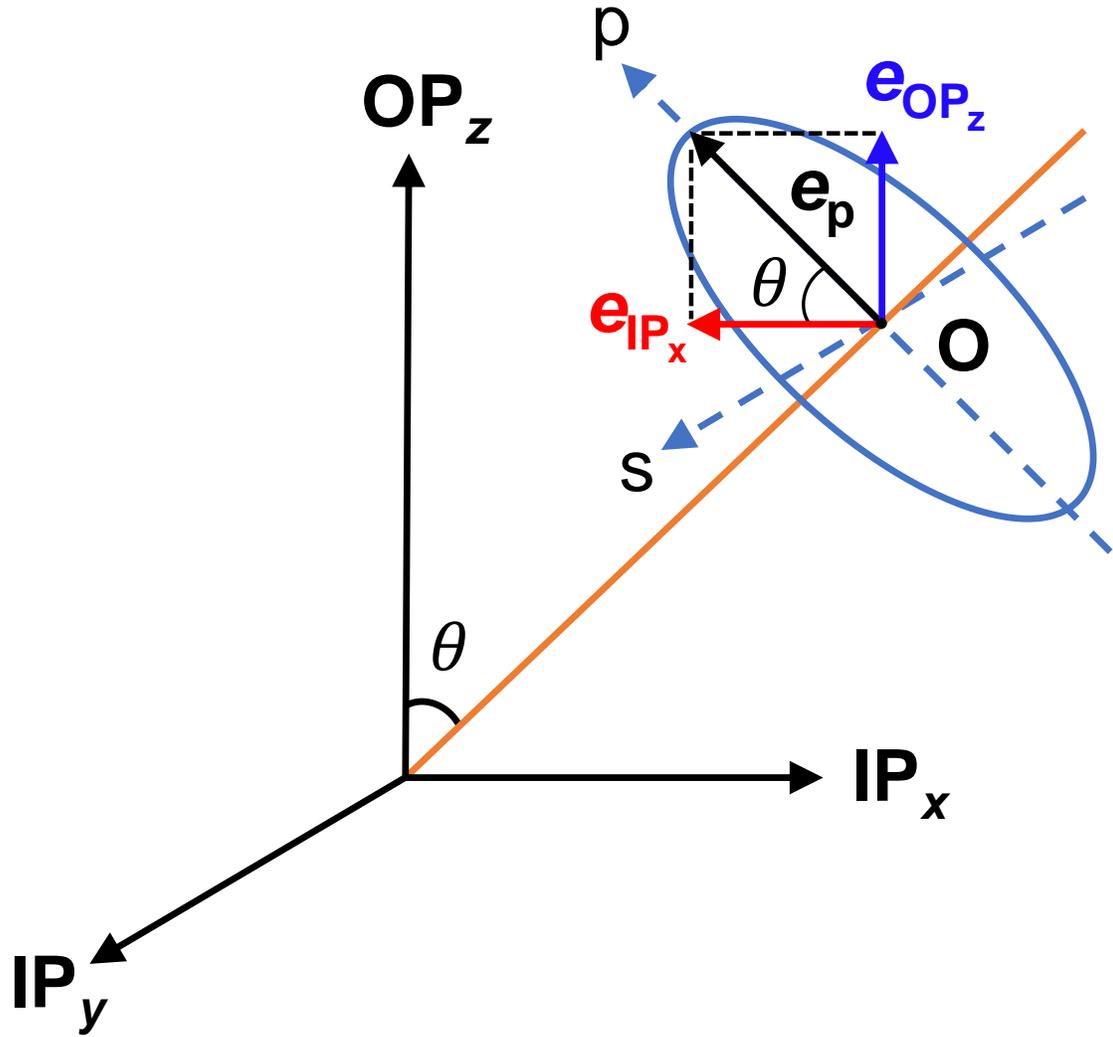

Fig. 10. Schematic of the decomposition of the p-polarized basis vector $e_p$ into the $IP_x$ and $OP_z$ basis vectors, $e_{IP_x}$ and $e_{OP_z}$, respectively. $\theta$ is the angle of incidence.

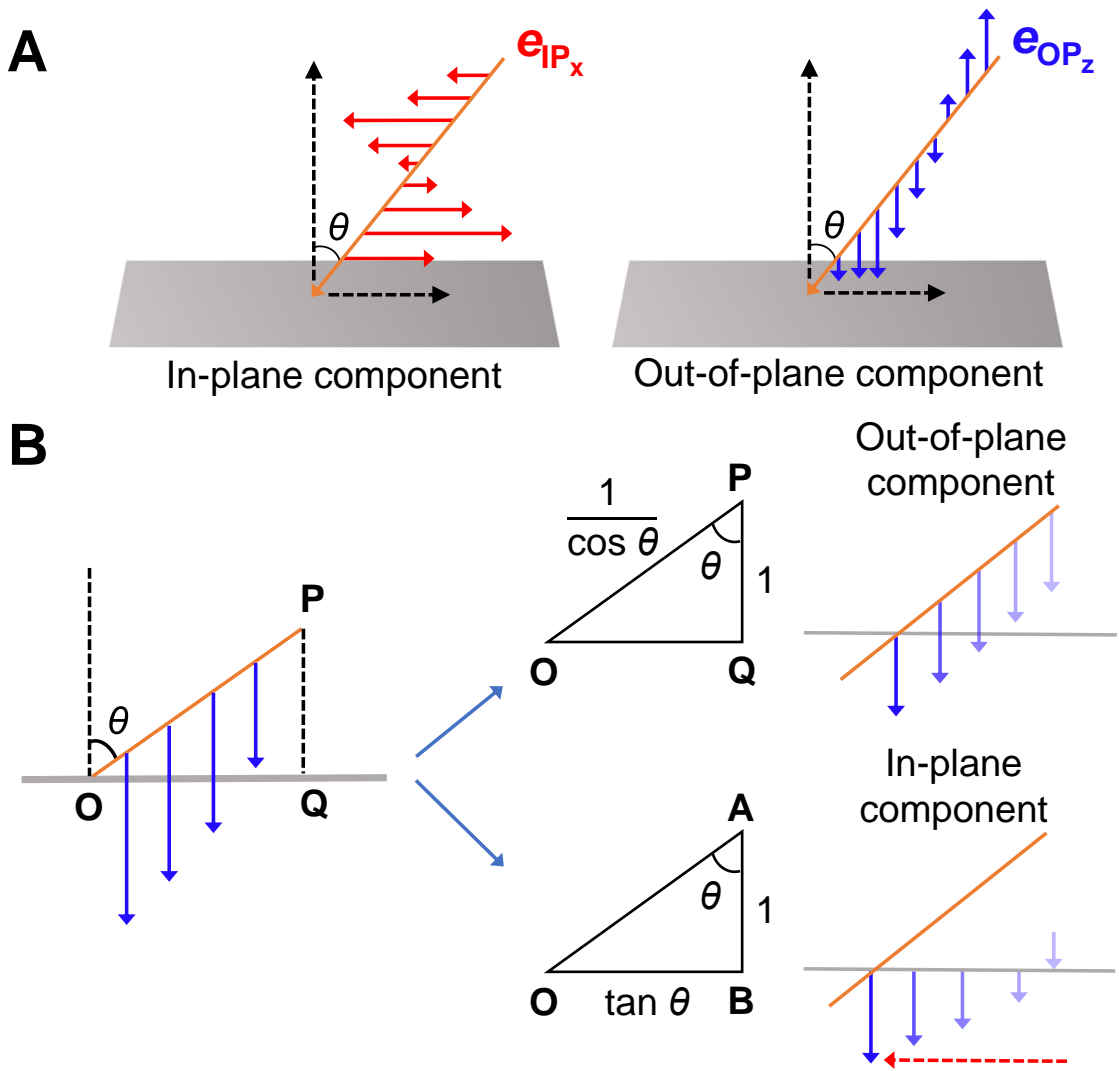

Fig. 11. (A) Schematic images showing the propagation of the $\cos\theta$ and $\sin\theta$ components of p-polarization. These components correspond to $|E_0|(\sin\phi)(\cos\theta)e_{IP_x} e^{i(k\cdot r-\omega t)}$ and $|E_0|(\sin\phi)(\sin\theta)e_{OP_z} e^{i(k\cdot r-\omega t)}$ in eq. 25 in the text, respectively. (B) Decomposition of the $\sin\theta$ component into out-of-plane and in-plane components by considering their propagation lengths induced by oblique incidence at the incident angle $\theta$.

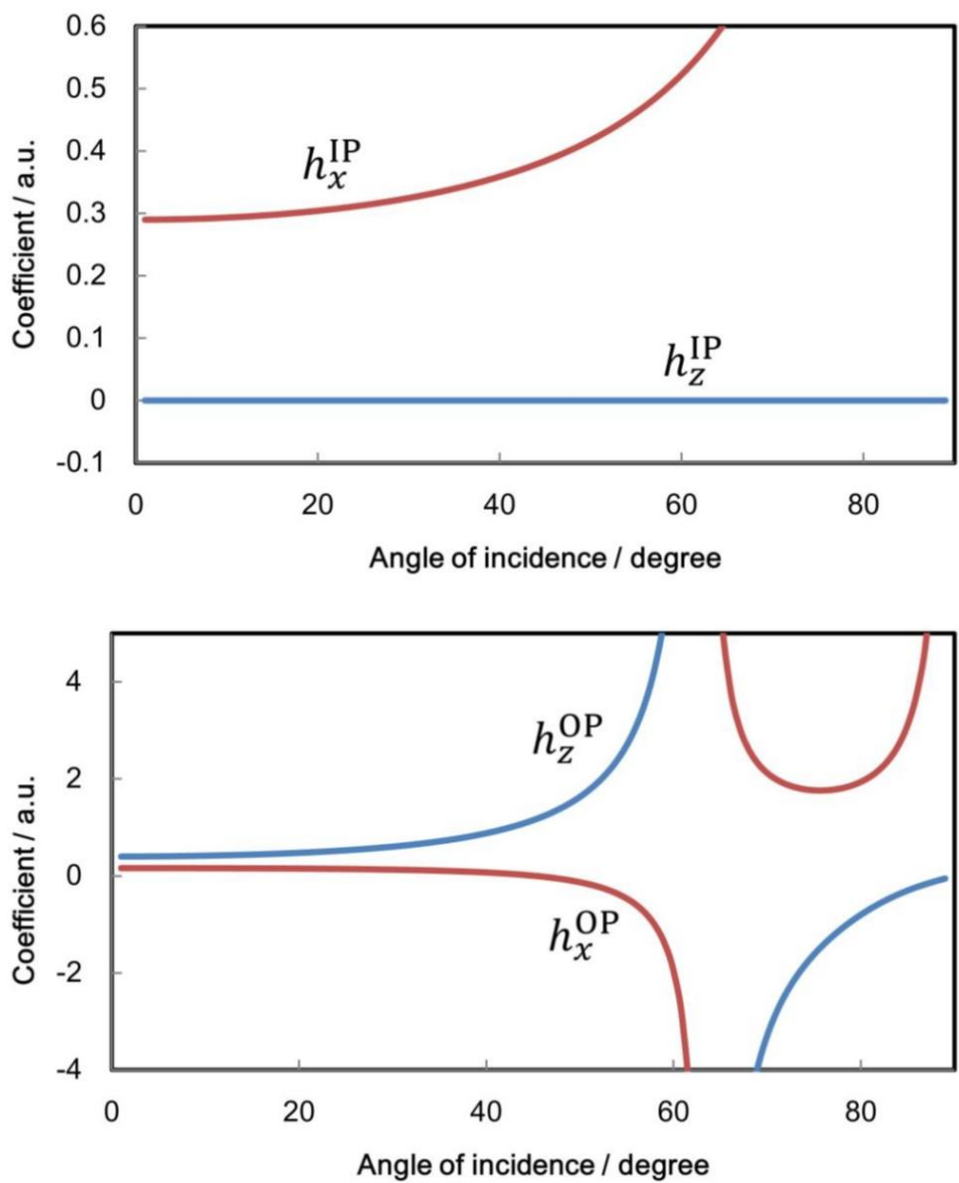

Fig. 12. Plots of the weighting coefficients, $h_{xy}^{IP}$, $h_z^{IP}$, $h_{xy}^{OP}$, and $h_x^{OP}$ in eqs. 33 and 34 in the text, against the angle of incidence ($1° \leq \theta \leq 89°$) calculated for a Si substrate. $h_x^{IP}$ and $h_x^{OP}$ are identical to $h_{xy}^{IP}$ and $h_{xy}^{OP}$ in the text of this article, respectively. Reprinted from ref.[32] with permission. Copyright 2019 American Chemical Society.

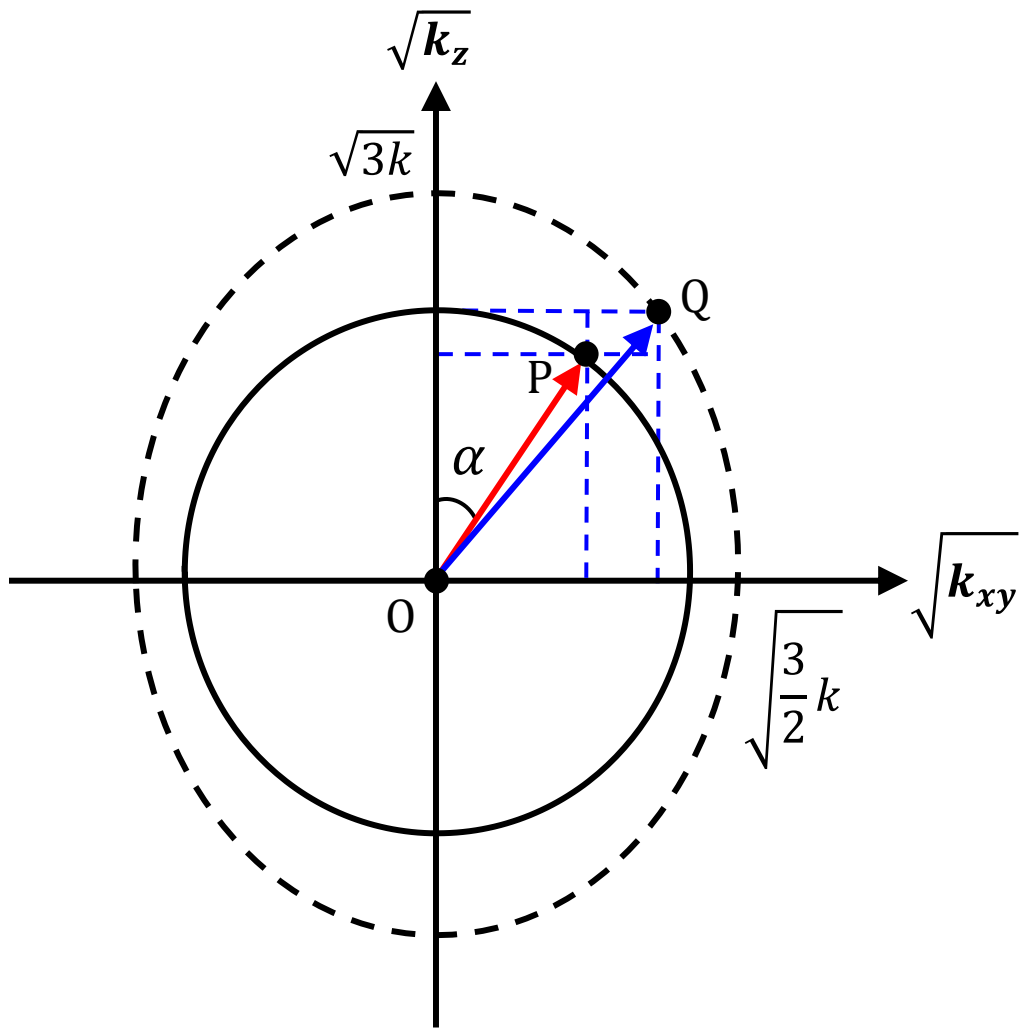

Fig. 13. Ellipsoidal representation of the relationship among the orientation angle $\alpha$, the optically isotropic extinction coefficient $k$, and its *x*- and *y*- (surface-parallel) and *z*- (surface-perpendicular) components, $k_{xy}$ and $k_z$, respectively, in a uniaxial system ($k_x = k_y = k_{xy}$). The ellipsoidal point Q represents the ideal point ($\sqrt{k_{xy}}, \sqrt{k_z}$). The norm |OP| represents the experimentally observed extinction coefficients.

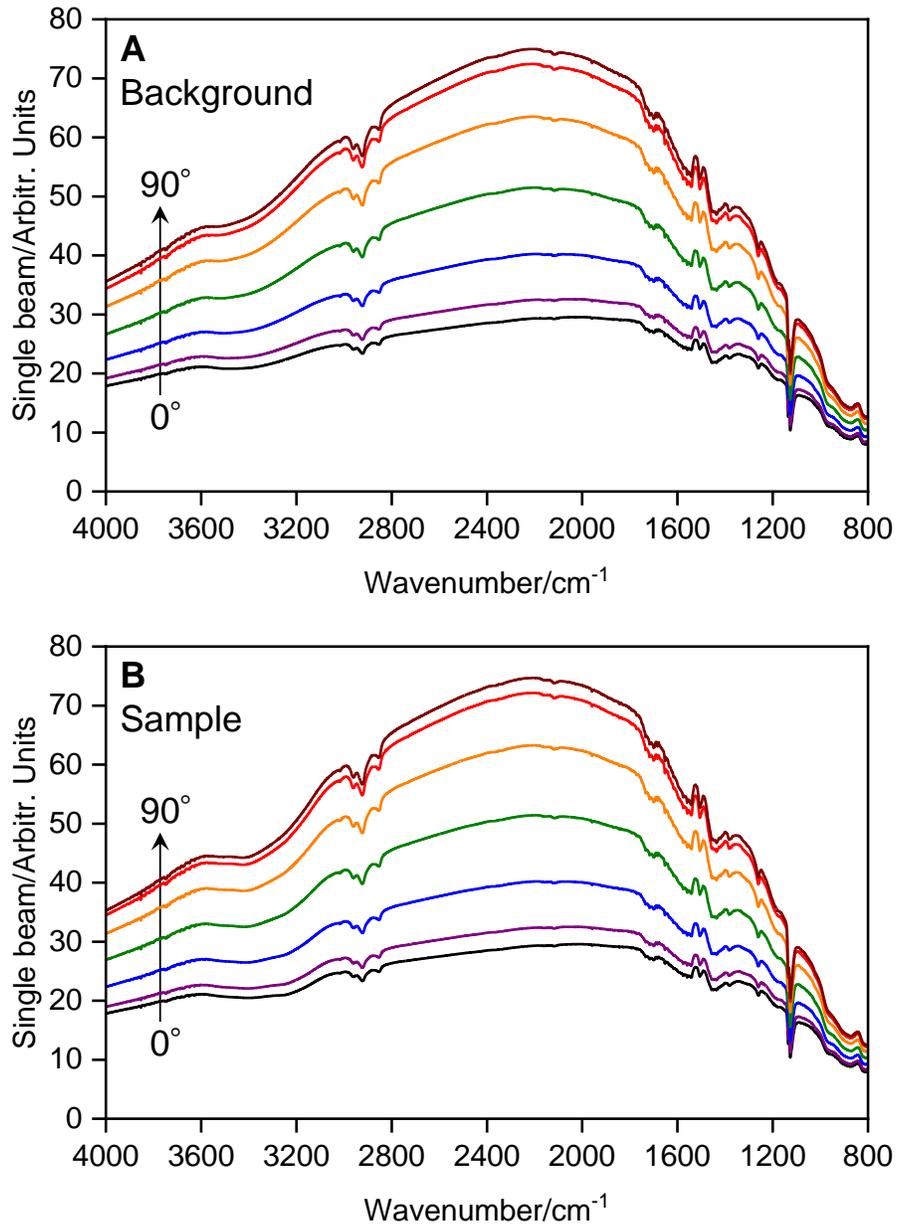

Fig. 14. Typical single-beam spectra measured at an angle of incidence of $\theta = 45°$ at various polarization angles from $\phi = 0°$ (s-polarization) to $\phi = 90°$ (p-polarization) in 15° steps before the IR-MAIRS analysis. (A) Background measurements using a bare Si substrate at 90 K, and (B) sample measurements for amorphous water on the Si substrate at 90 K formed by 32 min exposure of $H_2O$ (with 3.5 mol% HDO) at $2.2 \times 10^{-6}$ Pa.

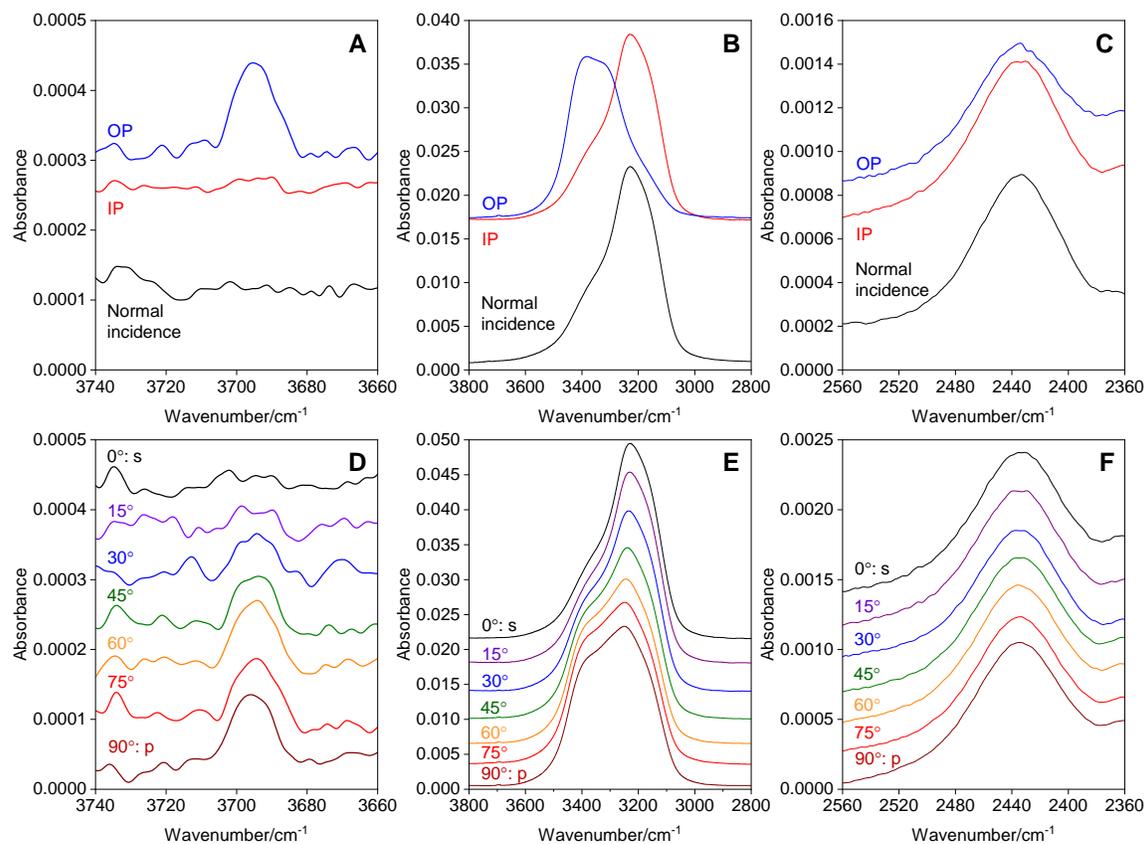

Fig. 15. In-plane (IP) and out-of-plane (OP) spectra of amorphous water on a Si substrate at 90 K. (A) Dangling OH stretching at 3740–3660 cm$^{-1}$. (B) OH stretching at 3800–2800 cm$^{-1}$. (C) Decoupled OD stretching at 2560–2360 cm$^{-1}$. The IP and OP spectra are calculated using the data in Fig. 14. Bottom black lines in (A), (B), and (C) show the normal-incidence IR spectrum ($\theta = 0°$, $\phi = 90°$). (D–F) Seven absorbance spectra calculated from the $j$th single-beam spectra in the background and sample measurements ($S_{\text{obs},j}^{\text{bg}}$ and $S_{\text{obs},j}^{\text{sam}}$, respectively) at each polarization angle of $\phi_j$ ($j = 1, 2, …, 7$), where $j = 1, 2, …, 7$ corresponding to $\phi = 0°$ (s-polarization), 15°, …, 90° (p-polarization), respectively, using the data in Fig. 14.

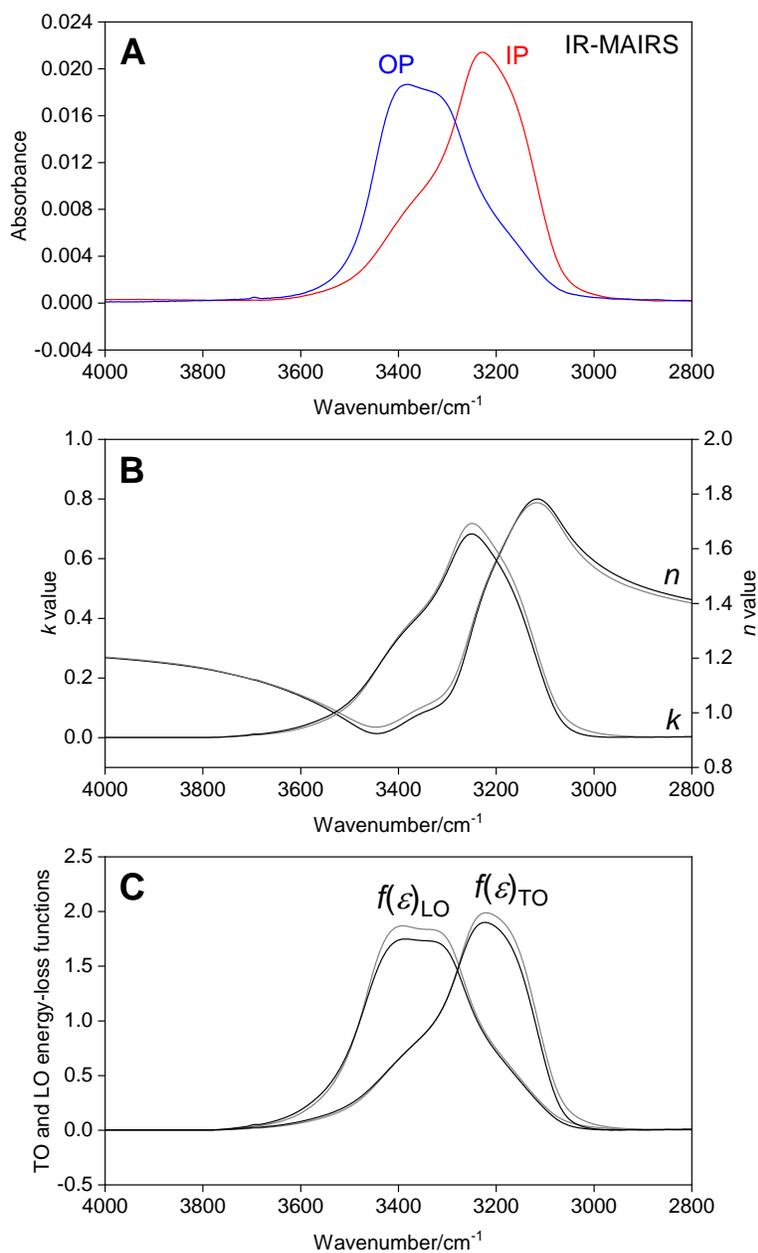

Fig. 16. Infrared spectra of vapor-deposited amorphous water at 90 K in the OH-stretching vibrational region at 4000–2800 cm$^{-1}$. (A) In-plane (IP) and out-of-plane (OP) spectra obtained by IR-MAIRS. (B) Literature values for the real ($n$) and imaginary ($k$) parts of the complex refractive index of amorphous water at 80 K (black) and 100 K (gray) from ref.[18]. (C) Transverse optic (TO) energy-loss function ($f(\varepsilon)_{TO}$) and longitudinal optic (LO) energy-loss functions ($f(\varepsilon)_{LO}$) calculated using the $n$ and $k$ values at 80 K (black) and 100 K (gray) from ref.[18]. The values for $f(\varepsilon)_{LO}$ are multiplied by three for clarity.

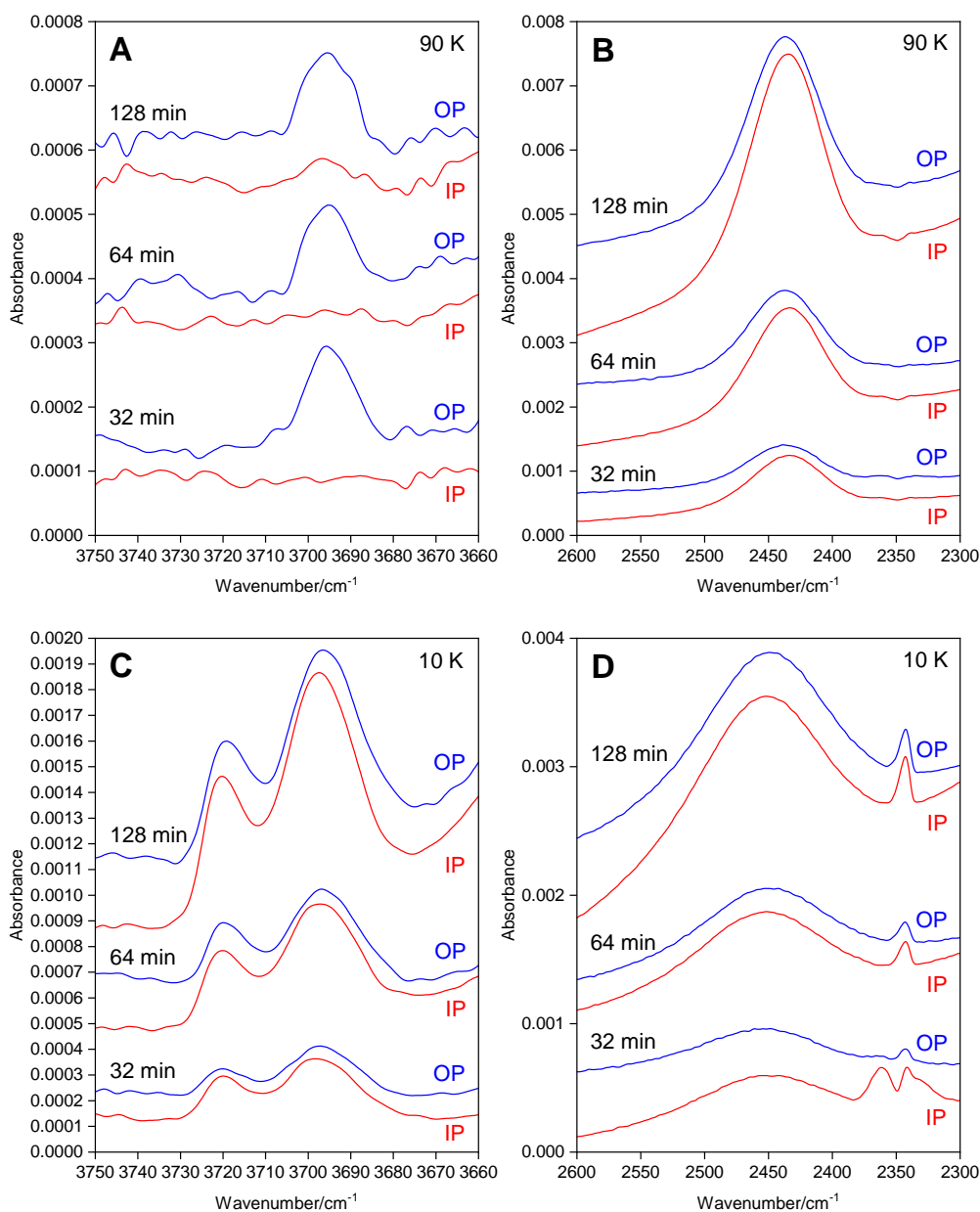

Fig. 17. In-plane (IP) and out-of-plane (OP) spectra of amorphous water on a Si substrate obtained by IR-MAIRS with respect to water exposure time. (A) Dangling OH stretching at 3750–3660 cm$^{-1}$ at 90 K, (B) decoupled OD stretching at 2600–2300 cm$^{-1}$ at 90 K, (C) dangling OH stretching at 3750–3660 cm$^{-1}$ at 10 K, and (D) decoupled OD stretching at 2600–2300 cm$^{-1}$ at 10 K. The pressure in the chamber was $2.2 \times 10^{-6}$ Pa during water exposure. The amorphous water samples were prepared using H$_2$O with 3.5 mol% HDO.

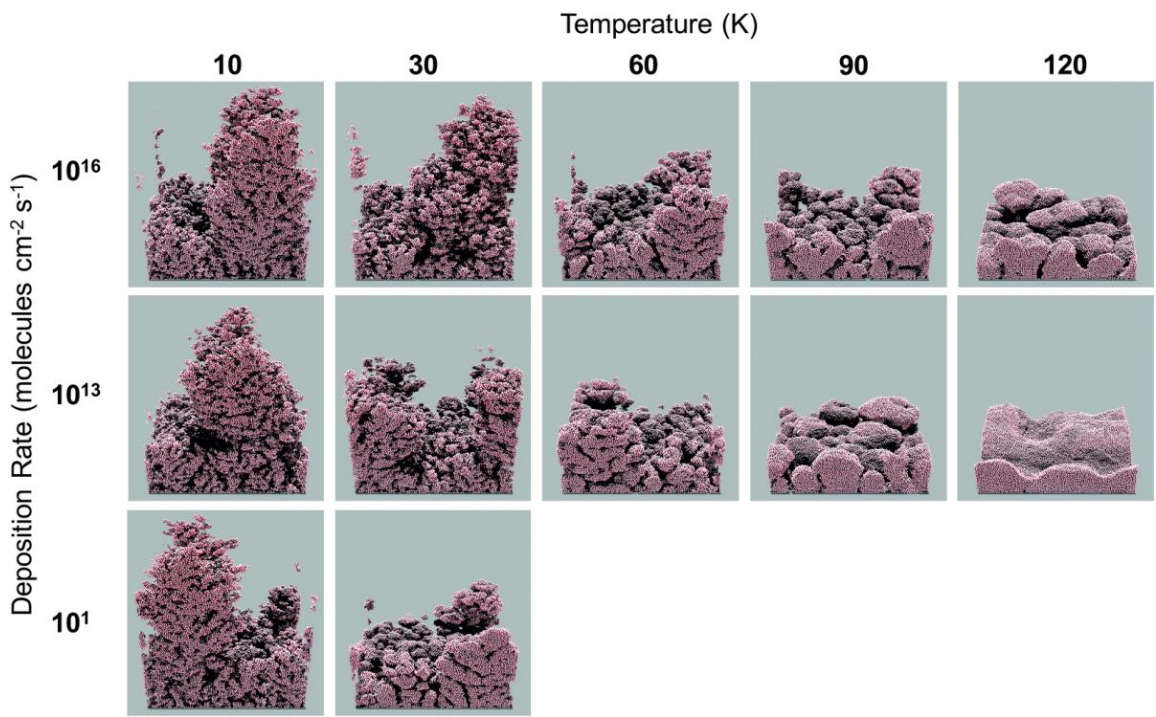

Fig. 18. Kinetic Monte Carlo simulations of water ice formed through background deposition at a range of rates ($10^1$, $10^{13}$, and $10^{16}$ molecules cm$^{-2}$ s$^{-1}$) and temperatures (10–120 K). Reproduced from ref.[82] with permission from the Royal Society of Chemistry.

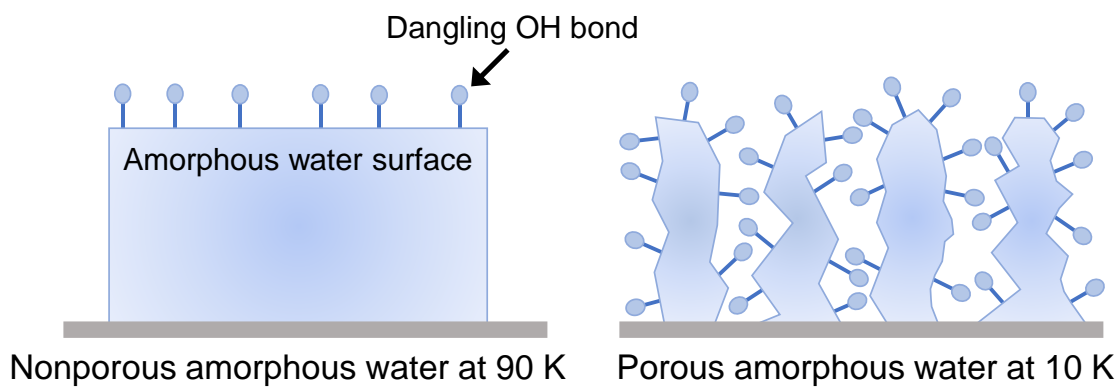

Fig. 19. Schematic illustrations of possible structures of vapor-deposited amorphous water at 90 K (left) and 10 K (right) estimated from IR-MAIRS measurements of dangling OH bonds (see also Fig. 17 A and C for the spectral data at 90 and 10 K, respectively).

Table 2. Typical peak wavenumber, height, width, and area of the decoupled-OD stretching band for vapor-deposited amorphous water at 90 K[a]

| Decoupled-OD stretching | Peak wavenumber [cm$^{-1}$] | Height | Width[b] [cm$^{-1}$] | Area [cm$^{-1}$] (2540–2380 cm$^{-1}$) |
|---|---|---|---|---|
| IP spectrum in IR-MAIRS | 2433 | 0.00057 | 61 | 0.0374 |
| Normal-incidence transmission | 2433 | 0.00058 | 61 | 0.0374 |

[a] The spectral data are shown in Fig. 15C.

[b] Full width at half-maximum height

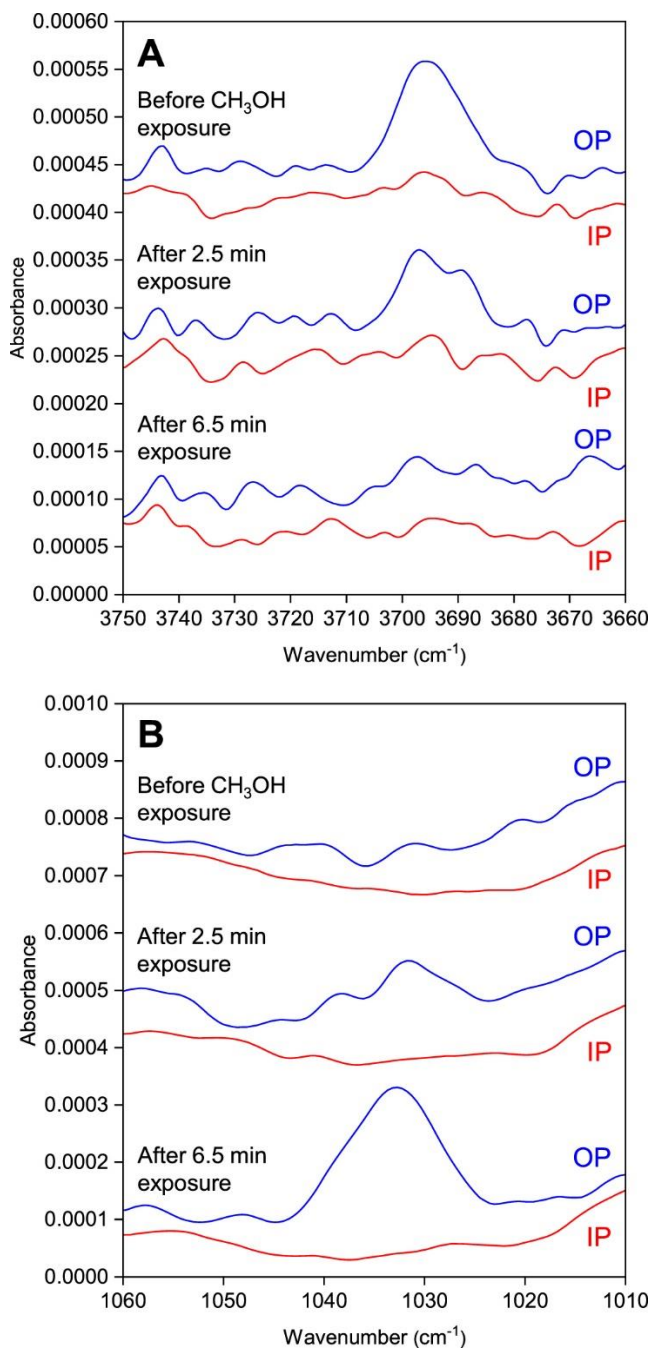

Fig. 20. In-plane (IP) and out-of-plane (OP) spectra of amorphous water on a Si substrate at 90 K with respect to $CH_3OH$ exposure time. (A) 3750–3660 cm$^{-1}$ for the dangling-OH peak and (B) 1060–1010 cm$^{-1}$ for the CO stretching band of $CH_3OH$. Amorphous water was prepared at 90 K by 32 min water exposure ($H_2O$ with 3.5 mol% HDO) at $2.2 \times 10^{-6}$ Pa before $CH_3OH$ exposure. The pressure in the chamber was $2.2 \times 10^{-8}$ Pa during $CH_3OH$ exposure. Reprinted from ref.[34]. Copyright 2021 the American Astronomical Society. Reproduced by permission.

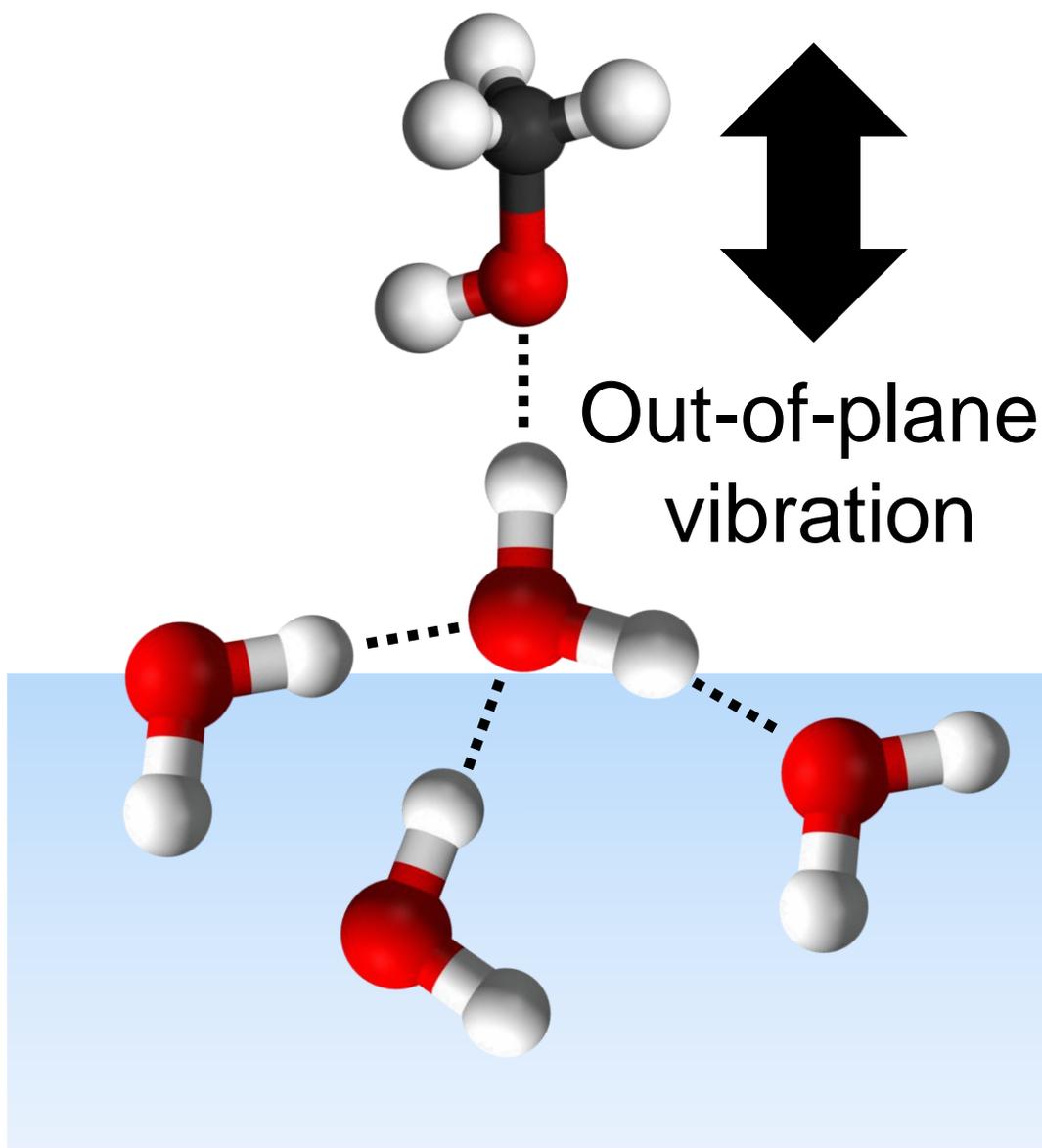

Fig. 21. Schematic of a possible configuration of a $CH_3OH$ molecule adsorbed on a three-coordinated dangling-OH bond on the surface of vapor-deposited amorphous water at 90 K.

Table 3. Summary of band strengths [cm molecule$^{-1}$] for water

| Species | Band strength (wavenumber range) |
| --- | --- |
| Dangling OH at 90 K (This study)[34] | $1.4 \pm 0.3 \times 10^{-17}$ (3710–3680) |
| Dangling OH (Calculation)[90] | $1.2 \times 10^{-17}$ (3700–3640) |
| Amorphous water at 80 K[18] | $2.9 \times 10^{-16}$ (4320–2956) |
| Crystalline ice at 267–273 K[a][91] | $2.3 \times 10^{-16}$ (3900–2700) |
| Liquid water at 295–298 K[a][92] | $1.5 \times 10^{-16}$ (3900–2700) |
| Monomer in solid CO at 10 K[95] | $1.1 \times 10^{-17}$ (3707)[b] |
| Monomer in solid N$_2$ at 10 K[95] | $7.8 \times 10^{-18}$ (3726)[b] |
| Monomer in solid O$_2$ at 10 K[95] | $3.3 \times 10^{-18}$ (3732)[b] |

[a] See ref[34] for details of the band strength calculations for crystalline ice and liquid water

[b] Asymmetric OH stretching vibration

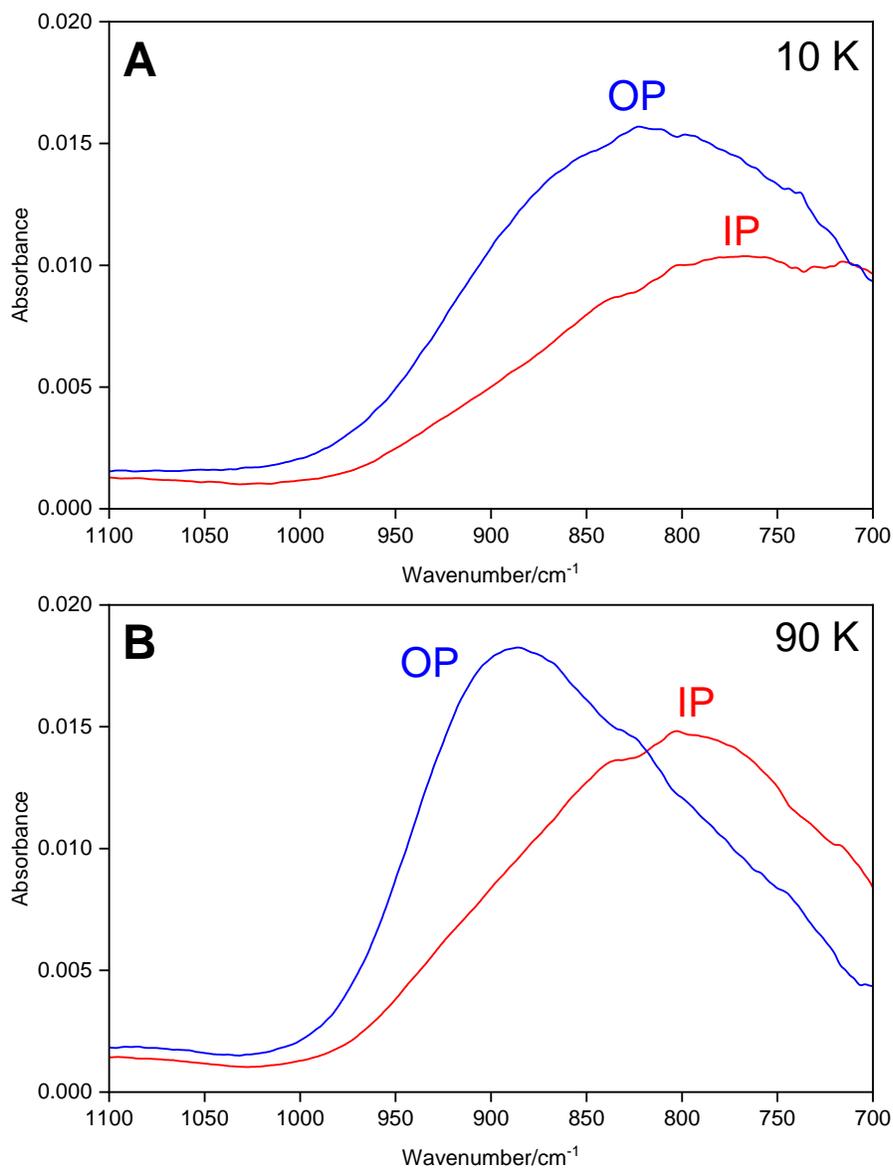

Fig. 22. In-plane (IP) and out-of-plane (OP) spectra of amorphous water on a Si substrate at (A) 10 K and (B) 90 K at 1100–700 cm$^{-1}$ for the libration band. Amorphous water samples were prepared at (A) 10 K and (B) 90 K by 128 min water exposure ($H_2O$ with 3.5 mol% HDO) at $2.2 \times 10^{-6}$ Pa.